*The optimal timing of reintroducing captive populations into the wild*

Richard T Melstrom[2,b]
melstrom@okstate.edu

Kehinde Rilwan Salau[3,c]
krsalau@math.arizona.edu

David W Shanafelt[1,4,a,d]
dshanafe@asu.edu

[1] Centre for Biodiversity Theory and Modelling. Theoretical and Experimental Ecology Station.
[2] Department of Agricultural Economics, Oklahoma State University
[3] Department of Mathematics, University of Arizona
[4] School of Life Sciences, Arizona State University

[a] Arizona State University, School of Life Sciences, PO Box 874601, Tempe, AZ 85287, USA
[b] Department of Agricultural Economics, Oklahoma State University, 317 Agricultural Hall, Stillwater, OK 74078, USA
[c] Department of Mathematics, The University of Arizona, 617 North Santa Rita Ave, Tucson, AZ 85721, USA
[d] Theoretical and Experimental Ecology Station, CNRS and Paul Sabatier University, 09200 Moulis, France

**Abstract**

We examine a conservation problem in which the recovery of an endangered species depends on a captive breeding and reintroduction program. The model is applied to the case of the black-footed ferret (*Mustela nigripes*), an endangered species in North America reliant on captive breeding for survival. The timing of reintroduction is an important concern in these programs as there is a tradeoff between the duration (and therefore the cost) of the captive breeding program and the period the population spends in recovery and in the wild. In this paper, we develop a stylized bioeconomic model to determine the optimal reintroduction time, in which the objective is to minimize the cost of reintroduction while providing a viably-sized population in the wild. Our control variable is the timing of reintroduction, which departs from a large body of work in bioeconomics that focuses on adjustable controls that directly affect the target population. Generally, we find it is optimal to reintroduce ferrets early in a reintroduction program, although this result is contingent on species interactions and provisioning services.

# 1. Introduction

Wildlife translocation and reintroduction is a popular but costly conservation tool. Over the last few decades, several hundred reintroduction programs have augmented existing or reestablished extirpated wildlife populations (Wolf et al. 1996). However, these programs are frequently criticized as expensive and ineffective (Ricciardi and Simberloff 2009; Kleiman, Stanley, and Beck 1994). Translocation typically follows a period of captive breeding, which requires substantial investments in facilities, fodder and genetic diversity maintenance, and many reintroductions fail to establish a sustainable population (Wolf et al. 1996).

One of the flagship species for reintroduction in North America is the black-footed ferret (*Mustela nigripes*). Originally prevalent across the Great Plains, mountain basins, and semi-arid grasslands of North America, the black-footed ferret became one of the most endangered species as a result of the sharp decline in its primary food source, prairie dogs, due to habitat conversion, disease, and poisoning (Miller, Reading, and Forrest 1996). In fact, between 1979 and 1981, the black-footed ferret was presumed extinct (USFWS 2013). In 1981, a remnant population was discovered in Meeteetse, Wyoming but after a series of disease outbreaks in the early 1980s all surviving ferrets were removed from the wild and placed in captive breeding programs. The ferret recovery strategy must combine captive breeding and reintroduction to achieve a population level that no longer warrants a listing of threatened or endangered under the U.S. Endangered Species Act (ESA). Reintroduction programs were initiated post-1990, with mixed success. Currently it is estimated that there is a minimum of 418 adult, breeding individuals among 11 reintroduction sites, with another 280 individuals in captivity (USFWS 2013).

This paper analyzes the optimal timing of translocation while accounting for the costs of managing captive and wild populations. We develop a bioeconomic model of captive rearing and reintroduction, including a wild prey population for the reintroduced individuals. We then use this model to determine the reintroduction time that minimizes total management costs, which include post-reintroduction costs associated with disease management, among other economic values. If the management objective is to minimize the costs of reintroduction subject to providing a viable long-term population in the wild, then a manager must consider the direct financial costs of managing the species in question, plus any indirect costs and benefits generated by the reintroduced species.

Existing studies of reintroduction programs tend to focus on post-reintroduction outcomes.[1] Population viability analyses are commonly used to account for stochastic events and identify minimum viable populations for endangered species (Beissinger and Westphal 1998). However, these analyses abstract from economic considerations. For black-footed ferrets, research has also focused on the influence of age structure (Klebanoff et al. 1991), habitat heterogeneity and space (Bevers et al. 1997), and uncertainty (Seal et al. 1989) in preparing for reintroductions. Thus, there is a critical need for research that incorporates economics into reintroduction problems.

Bioeconomic models that integrate economic and biophysical components are often used to show that failure to consider relevant costs and benefits in resource conservation leads to inefficient outcomes (Clark 1973; Clark and Munro 1976; Clark and Kirkwood 1986; Wilen and Brown 1986; Mesterton-Gibbons 1987, 1988; Crocker and Tschirhart 1992; Brock and

---

[1] Reintroduction programs for endangered species often depend on captive breeding programs that risk creating genetic bottlenecks. Thus, much of the literature on the biology of managing captive populations focuses on genetic variability (Seal et al. 1989).

Xepapadeus 2002). Among other topics, these models have been used to analyze fisheries (Gordon 1954; Hilborn 1976; Sanchirico and Wilen 1999), biological invasions (Leung et al. 2002; Horan and Fenichel 2007; Fenichel, Horan, and Hickling 2010; Epanchin-Niell and Hastings 2010; Homans and Horie 2011), and endangered species (Alexander 2000; Melstrom and Horan 2014). Several bioeconomic papers on the subject of endangered species examine reintroduction problems. Rondeau (2001) was among the first to examine the management of a reintroduced population with a bioeconomic model, while Horan and Melstrom (2011) include the possibility of reintroducing captive animals if the wild population goes extinct. Canessa et al. (2015) examine the issue of reintroduction under uncertainty.

This paper contributes to bioeconomic research by explicitly examining the timing of translocation for captive breeding and reintroduction programs. We build on prior work by using optimal control theory to determine the solution to an optimization problem in which time, rather than a stock or harvest variable, is the control. Very few studies consider time as the control (though see Amit (1986)). Although conceptually similar, the classic Faustmann rotation problem and its extensions consider a single period of net benefits in the optimization process (Conrad 1999; Newman, Gilbert, and Hyde 1985; Loisel 2011; Olschewski and Benitez 2010; van Kooten, Binkley, and Delcourt 1995). This is in contrast with our problem, in which benefits or costs are accrued before and after reintroduction. Researchers may therefore find the methods in this paper useful in solving other endogenous time problems. The solution to black-footed ferret application implies that translocation and reintroduction should either take place immediately or be delayed indefinitely, depending on economic values and the biology of the captive population. We confirm this solution using brute-force numerical optimization.

## 2. Background

Black-footed ferrets are one of the most endangered species in the world. The species, a type of mustelid native to the prairies and mountain grasslands of North America, is famous for its $20^{th}$ century decline and eventual status as extinct-in-the-wild.

Originally listed as endangered in 1967 (Udall 1967) and 1970 (Russell 1970), the black-footed ferret was grandfathered into the Endangered Species Act of 1973. Its historical range comprised approximately 100 million acres (40.5 million hectares) of intermountain and prairie grasslands stretching from Canada to Mexico wherever prairie dogs occurred (Hillman and Clark 1980; Anderson et al. 1986). Estimates of historic ferret populations range between half a million to one million individuals (Anderson et al. 1986). However, starting at the end of the nineteenth century, the ferret population was gradually devastated by declines in prairie dogs due to habitat conversion, poisoning, and disease (USFWS 2013). By 1964 ferrets were classified as "extremely rare" (Henderson, Springer, and Adrian 1969) and presumed extinct in 1979 (USFWS 2013). In 1981 a small population was discovered in Meeteetse, Wyoming (Clark 1986). This population declined to 15 individuals (equivalent of 7 genetically-distinct founders) before it was placed in captivity to support a captive breeding and reintroduction program (USFWS 2013). No wild populations have been found since the removal of the Meeteetse population - it is likely that all current black-footed ferrets are descended from those 15 individuals (Hanebury and Biggins 2006).

Since 1991 there have been 20 black-footed ferret reintroduction projects, which have had mixed success (USFWS 2013).  As of 2011, the U.S. Fish and Wildlife Service program had produced more than 7,000 ferrets in captivity, of which more than 2,600 had been translocated to the wild.  Currently, population estimates suggest there are less than 500 breeding adult ferrets in the wild, although the number of known, documented wild ferrets is much less (USFWS 2013). The scale of difference between the number of produced ferrets and the size of the existing wild population suggests there may be inefficiencies in the reintroduction program.

To delist black-footed ferrets as an endangered species, the following criteria must be met for a period of at least three years (USFWS 2013):

- Maintain a captive breeding population of at least 280 individuals (at least 105 males and 107 females) across at least three captive breeding facilities
- Establish a free-range population of at least 3,000 breeding individuals distributed across no less than 30 populations (at least one population in 9 of the 12 states within the historical range)
- Each population must have no fewer than 30 breeding individuals (at least 10 populations must have 100+ breeding individuals); at least five populations must contain Gunnison's and white-tailed prairie dog colonies
- Maintain approximately 200,000 hectares of occupied prairie dog habitat at black-footed ferret reintroduction sites, including plans for plague management and prairie dog conservation

To meet these criteria, managers must establish several dozen black-footed ferret reintroduction sites with viable populations, which requires knowledge about the dynamics of ferrets and prairie dogs.  Managers must be cognizant of the stock of prairie dogs prior to translocation as well as the potential ferret carrying capacity.  If the number of ferrets exceeds what can be supported by the stock of prairie dogs, the prey will be overexploited and the ferret population will crash.

Implicitly, the delisting criteria must be satisfied without exceeding the funding allocated to the national reintroduction project.  This means that managers are pursuing both ecological and economic goals.  Costs are incurred in raising ferrets in captivity and managing prairie dogs in the wild (pre-introduction), and managing both prairie dogs and ferrets in the wild (post-introduction).  Managers need to identify a reintroduction strategy that establishes viable populations in a least-cost manner.

**3. Ecology**

This section presents a basic, idealized population model of black-footed ferrets and prairie dogs. Black-footed ferrets are obligate predators, with prairie dogs comprising over 90% of their diet (Sheets, Linder, and Dahlgren 1972; Campbell et al. 1987).  The ferrets' historic range coincides with three species of prairie dog (white-tailed, black-tailed, and Gunnison's), though no study indicates a preference for any species.  Ferrets also use prairie dog burrows for shelter (Biggins 2006; Hillman 1968).  Much is known about the demography of ferrets and prairie dogs (USFWS 2013; Holmes 2008; Miller, Reading, and Forrest 1996; Hoogland et al. 1987; Knowles 1982; Reading 1993) but relatively little work has modeled these two species jointly as a predator-prey system (though see Bevers et al. (1997) and Klebanoff et al. (1991)).

Consider a potential black-footed ferret reintroduction site, e.g. a single patch that can support black-footed ferrets and prairie dogs. We denote these two populations as $p$ and $f$ for prairie dogs and ferrets, respectively. This predator-prey system is defined by the following system of equations:

$$[1] \quad \frac{dp}{dt} = g(p(t))p(t) - h(p(t))f(t)$$

$$[2] \quad \frac{df}{dt} = \beta h(p(t))f(t) - m(f(t))$$

where $g(p(t))$ is the growth rate of prairie dogs, $h(p(t))$ is the consumption of prairie dogs per ferret (the functional response), $\beta$ is the conversion efficiency of prairie dog to ferret biomass, and $m(f(t))$ is mortality of ferrets at time $t$. Several studies show that prairie dogs exhibit logistic growth (Miller, Reading, and Forrest 1996; Reading 1993; Garrett, Hoogland, and Franklin 1982; Knowles 1982). This implies that $g(p(t)) = r(1 - p(t)/K)$ where the prairie dog intrinsic growth rate and carrying capacity are given by $r$ and $K$, respectively. Bevers et al. (1997) suggest that ferrets also experience logistic growth, though other studies attribute this phenomena to a functional response to prey populations rather than logistic growth in the predator species (Klebanoff et al. 1991). The exact nature of the functional response of ferrets to prairie dogs is unclear. In the application, we consider three different predator response functions (Holling 1959).[2]

Modeling ferret-prairie dog interactions as the system of equations [1]-[2] requires several assumptions about ferret-prairie dog dynamics. First, no dispersal occurs for either species. Both ferrets and prairie dogs are territorial species, and young often disperse from their natal colonies (Miller, Reading, and Forrest 1996; USFWS 2013; Hoogland 2013; Garrett and Franklin 1988). By considering a single, representative reintroduction site we assume there is sufficient "space" for individuals to disperse and form new colonies within the boundaries of the site. With few exceptions, ferret reintroduction sites are spatially distinct such that no natural dispersal occurs between them (USFWS 2013). Second, we aggregate all age and sex classes into a single compartment for each species. This assumption is common in the ferret modeling literature (though see Klebanoff et al. (1991) for an example of an age-structured model). Third, the speed of dynamics is the same for ferrets and prairie dogs. Finally, we do not explicitly model dynamics associated with disease or higher-level trophic predators; these issues are dealt with implicitly through existing parameters (e.g. ferret mortality rate).

---

[2] For a detailed description of the different predator response functions, see Gotelli (1995).

## 4. Bioeconomics

The recovery and delisting objectives faced by managers require growing and translocating captive populations to the wild at minimal cost. This implies the following cost minimization problem:

[3]
$$\min_{\tau} \int_0^{\tau} \left[ c_c\left(f_c(t)\right) + c_p\left(p_c(t)\right) \right] e^{-\delta t} dt + \int_{\tau}^{T} \left[ c_w\left(f_w(t)\right) + c_p\left(p_w(t)\right) \right] e^{-\delta t} dt$$
$$+ \varphi\left(T, f_w(T), p_w(T)\right)$$

subject to:
$$\frac{df_c}{dt}, \frac{df_w}{dt}, \frac{dp_c}{dt}, \frac{dp_w}{dt}$$
$$f_c(0), p_c(0)$$
$$f_c(\tau) = f_w(\tau)$$
$$p_c(\tau) = p_w(\tau)$$

where managers choose the time, $\tau$, to reintroduce a captive population of ferrets to the wild. The stock of ferrets in captivity is denoted by $f_c$, ferrets in the wild by $f_w$. The stocks of prairie dogs in the wild pre- and post- reintroduction are given by $p_c$ and $p_w$ respectively.

The planning horizon ends at time *T*, an exogenous parameter. Other bioeconomic models of wildlife conservation similarly assume that management occurs within a fixed interval (Haight 1995; Spring et al. 2001; Nalle et al. 2004). Management of imperiled species is often viewed as temporary, lasting until it is clear extinction risks will not occur. Consequently, endangered species recovery and habitat conservation plans typically have expected end dates (Suckling 2006; Bernazzani, Bradley, and Opperman 2012).[3] We assume that by time *T in situ* management is no longer required in the wild because problems with habitat and disease, which are beyond the scope of our model, will have been solved.

The scrap "cost" ($\varphi$) is associated with costs outside of direct management after the end of the conservation plan (e.g. additional monitoring and indirect costs such as prairie dog damages to farmland). The discount rate is given by δ.

In addition, we evaluate the minimum and maximum numbers of ferrets and prairie dogs that may be present at the point of introduction ($f_{min} \leq f_c(\tau) \leq f_{max}$ and $p_{min} \leq p_c(\tau) \leq p_{max}$) as biological criteria for reintroduction. They are the combinations of ferrets and prairie dogs (initial conditions) that result in a positive, long-run coexistence of the two species. These criteria are not considered explicitly in the optimization of [3], but are instead evaluated *ex poste* as a check for suitable biological conditions necessary for reintroduction. We further assume that during the pre-introduction stage no ferrets are present in the reintroduction site ($f_w = 0$ if $t < \tau$) and that all captive ferrets are released into the wild at time $\tau$ ($f_c = 0$ if $t > \tau$).

---

[3] Recent research shows that recovery strategies could be improved if the time duration was chosen optimally (Lampert et al., 2014).

There are several different cost functions. Viable ferret habitat must be populated with sufficient prairie dogs, which requires prairie dog monitoring, vaccinations and other general site management costs (USFWS 2013). Prairie dogs are sometimes thought of as pests, disturbing grazing sites for cattle. Furthermore, ferrets must be bred in captivity, which incurs food and maintenance costs, vaccinations, and training (USFWS 2013; Biggins et al. 1999). Ferret costs in captivity are captured by $c_c(f_c(t))$, with concurrent prairie dog monitoring costs of $c_p(p_c(t))$ appearing in the first term of [3]. After introduction, managers continue to expend resources associated with monitoring prairie dogs as well as wild ferrets (USFWS 2013; Holmes 2008). Prairie dogs must be monitored before and after ferret reintroduction, and because they are often viewed as a nuisance to productive rangelands, black-footed ferrets provide regulating ecosystem services through prairie dog control. These costs are found in the second term of [3] ($c_p(p_w(t))$ and $c_w(f_w(t))$ respectively). The cost function for prairie dogs can also be viewed as a damage function for variable rangeland.

In order to solve for the optimal reintroduction time, we adopt a first-principles approach following Conrad and Clark (1987) and Clark (2010). In optimal control theory, we often take the necessary first order condition(s), and adjoint and transversality equations as given. In fact they are derived by equating the change in the Lagrangian for a small change in the control to zero. We write the current-value Lagrangian describing the total costs of managing ferrets and prairie dogs in perpetuity as:

[4] $L(f_c, f_w, p_c, p_w, \tau, T) =$

$$\int_0^\tau \left[ \left[ c_c(f_c(t)) + c_p(p_c(t)) \right] e^{-\delta t} + \lambda_1(t) \frac{df_c}{dt} + \mu_1(t) \frac{dp_c}{dt} + \frac{d\lambda_1}{dt} f_c(t) + \frac{d\mu_1(t)}{dt} p_c(t) \right] dt$$

$$+ \left[ \lambda_1(0) f_c(0) + \mu_1(0) p_c(0) \right] - \left[ \lambda_1(\tau) f_c(\tau) + \mu_1(\tau) p_c(\tau) \right]$$

$$+ \int_\tau^T \left[ \left[ c_w(f_w(t)) + c_p(p_w(t)) \right] e^{-\delta t} + \lambda_2(t) \frac{df_w}{dt} + \mu_2(t) \frac{dp_w}{dt} + \frac{d\lambda_2}{dt} f_w(t) + \frac{d\mu_2}{dt} p_w(t) \right] dt$$

$$+ \left[ \lambda_2(\tau) f_w(\tau) + \mu_2(\tau) p_w(\tau) \right] - \left[ \lambda_2(T) f_w(T) + \mu_2(T) p_w(T) \right] + \varphi(f_w(T), p_w(T))$$

where $\lambda_1$ ($\mu_1$) and $\lambda_2$ ($\mu_2$) are the shadow values of an extra unit of ferret (prairie dog) biomass pre- and post- introduction respectively.

We use [4] together with the biological constraints in [3] to define the necessary conditions to solve for the optimal reintroduction time, $\tau^*$. Derivations of similar problems exists in Amit (1986), Kamien and Schwartz (1991), Dixit and Pindyck (1994), and Lenhart and Workman (2007).

At the optimal reintroduction time the difference between the Lagrangian evaluated at a candidate reintroduction time ($\tau$) and a candidate reintroduction time plus a small perturbation ($\tau + \varepsilon$) goes to zero (Conrad and Clark 1987; Clark 2010). This is equivalent to taking a partial derivative of the Lagrangian with respect to the reintroduction time (Lenhart and Workman 2007). Thus in order for $\tau^*$ to be an optimal reintroduction time, the following hold:

$$[5] \quad \Delta L = \frac{\partial L}{\partial \tau} + \left[\frac{\partial L}{\partial f_c} + \frac{\partial L}{\partial p_c}\right] + \left[\frac{\partial L}{\partial f_w} + \frac{\partial L}{\partial p_w}\right] = 0$$

That is, the change in the Lagrangian over the entire time horizon is equal to zero. Equation [5] forms the basis for deriving the traditional necessary first order conditions, adjoint equations, and transversality conditions in resource economics (Clark 2010; Conrad and Clark 1987).

The first term on the right-hand side of [5] is the direct change in the Lagrangian for a small change in the reintroduction time. This forms the necessary first order condition for cost minimization. The bracketed terms reflect the secondary effects that a small change in the reintroduction time has on the Lagrangian. Changes to the reintroduction time - an input to the biological model - will perturb ferret and prairie dog abundances and by extension, the Lagrangian. From [5] we can derive the following conditions that must hold for an optimal $\tau$ (see Supplemental Material A for details):

$$[6] \quad c_c\left(f_c(\tau)\right) = c_w\left(f_w(\tau)\right)$$

$$[7] \quad \frac{d\lambda_2}{dt} = -\frac{\partial}{\partial f_w}\left[c_w\left(f_w(t)\right) + \lambda_2(t)\frac{df_w}{dt} + \mu_2(t)\frac{dp_w}{dt}\right]$$

$$[8] \quad \frac{d\mu_2}{dt} = -\frac{\partial}{\partial p_w}\left[c_p\left(p_w(t)\right) + \lambda_2(t)\frac{df_w}{dt} + \mu_2(t)\frac{dp_w}{dt}\right]$$

$$[9] \quad \lambda_2(\tau) = 0$$
$$[10] \quad \mu_2(\tau) = 0$$

$$[11] \quad \lambda_2(T) = \frac{\partial}{\partial f_w}\varphi\left(f_w(T), p_w(T)\right)$$

$$[12] \quad \mu_2(T) = \frac{\partial}{\partial p_w}\varphi\left(f_w(T), p_w(T)\right)$$

Condition [6] forms the necessary first order condition for optimality. It says that at the time of reintroduction, $\tau^*$, the costs of managing ferrets in captivity must equal the costs of managing ferrets in the wild.[4] In other words, managers should only consider translocating ferrets from captivity into the wild when they are financially indifferent between continuing to nurture ferrets in a controlled environment and monitoring free roaming ferrets. This is a necessary but not a sufficient condition to determine exact time of reintroduction.[5]

---

[4] As in other studies of interacting species (Mesterton-Gibbons, 1988; Wilen and Brown, 1986), intuitively we would expect condition [6] to include prairie dogs. Prairie dogs drop out of the solution by the nature of the cost functions and the general setup of the problem. We would expect the cost functions of managing prairie dogs to be the same pre- and post-reintroduction. Since we choose the reintroduction time such that the change in the Lagrangian is zero, as the change in the candidate reintroduction time goes to zero the populations of prairie dogs on either side of the reintroduction times are equivalent. When substituted into their respective cost functions, they cancel each other out.

It is interesting to compare condition [6] with the solution to analogous resource problems; namely, timber management. For a stand of timber, the optimal single rotation requires solving for the length of time *t* where the marginal value of allowing the stand to grow equals the marginal cost of waiting to cut (Conrad 1999). Here, the marginal value of allowing the ferrets to grow in captivity is the cost savings of delaying reintroduction and spending on wild ferret management, and the marginal cost of delaying reintroduction is the spending required to support the captive breeding program. However, an important difference between our problem and a rotation problem is that ferret management suffers costs before and after reintroduction, and timber management collects benefits only at the time the stand is felled.

Conditions [7] and [8] describe the trajectory of the shadow values of ferret and prairie dogs post-reintroduction (adjoint equations). These shadow values measure the costs of an additional animal on managing ferrets and prairie dogs into the future. Conditions [9]-[12] form initial and terminal time (transversality) conditions for the shadow values in the post-reintroduction period of the problem. There are no adjoint equations from the pre-introduction part of the problem because the first part in the right hand side of [4] is, in essence, not controlled by management until the point of reintroduction. The solution only uses information about the system when a control is applied to the system - in our case, when ferrets are reintroduced into the wild (Clark 2010; Conrad 1999; Conrad and Clark 1987; Lenhart and Workman 2007).

It is worth reiterating that an optimal candidate reintroduction time must satisfy the full suite of conditions in equations [6]-[12] - the necessary first order condition alone is not sufficient. However, the system of equations in [6]-[12] generally precludes the possibility of an interior, *analytical* closed-form solution. The system is over-determined because an optimal $\tau^*$ must satisfy the necessary first order conditions and both initial *and* terminal boundary conditions - a unique solution that is difficult to satisfy for general suites of parameters (Cronin 2007). In order to overcome this, we develop a numerical method that explicitly accounts for the

---

[5] If we do not want to require that $\tau^*$ lies strictly between the initial time and final time, then a more complete set of necessary conditions include:

[6]    $c_c(f_c(\tau^*)) = c_w(f_w(\tau^*))$    if    $0 < \tau^* < T$

[6*]   $c_c(f_c(\tau^*)) \leq c_w(f_w(\tau^*))$   if    $0 < \tau^* = T$

[6**]  $c_c(f_c(\tau^*)) \geq c_w(f_w(\tau^*))$   if    $0 = \tau^* < T$

The additional conditions have an intuitive interpretation. Condition [6*] reflects scenarios in which it is always cost effective to manage ferrets in captivity. If the introduction time equals the initial time, then we have a scenario reflected in [6**] - it is cost effective to immediately release ferrets into the wild. Considering the realities of ferret management as well as the formulaic setup and constraints of [3], it is unlikely that conditions [6*] and [6**] reflect actual preferred choices even though they may be optimal. Condition [6*] requires the captive ferret population grow to its steady state (in the wild) amount while, simultaneously, the prairie dog population attains its equilibrium value (conditional on there being wild ferrets) - a scenario that can be attained in simulations but would be difficult to achieve in practice. Condition [6**] is improbable because newborn captive ferrets often require time to learn the skills necessary to survive in the wild (Hutchins et al. 1996; USFWS, 2013).

population dynamics of ferrets and prairie dogs and the changing costs associated with altering the reintroduction time. This is discussed in the next section.

## 5. Application

*5.1. Biology*
We now specify functional forms for the ferret-prairie dog system. A detailed understanding of the biological system allows us to characterize the conditions required for coexistence and map the quantities of each species required to establish a long-term population. First, we assume that ferrets in captivity grow at a constant growth rate. Second, as discussed in the ecology section, growth of prairie dogs is modeled as a logistic function. For simplicity, the mortality of ferrets is assumed to occur at a fixed rate, so that $m(f(t)) = df(t)$. The choice of functional response is nontrivial, so we examine three functional forms commonly used in the ecology literature. We first characterize the functional response as a type I response – where the predation rate remains constant regardless of the number of prey (prairie dogs) and predators (ferrets). This implies

[1*] $$\frac{dp}{dt} = rp(t)\left(1 - \frac{p(t)}{K}\right) - \alpha p(t) f(t)$$

[2*] $$\frac{df}{dt} = \alpha \beta p(t) f(t) - df(t)$$

where $\alpha$ is a measure of predation efficiency. By contrast, a type II predator response transforms equations [1*] and [2*] into:

[1**] $$\frac{dp}{dt} = rp(t)\left(1 - \frac{p(t)}{K}\right) - \left(\frac{\kappa p(t)}{D + p(t)}\right) f(t)$$

[2**] $$\frac{df}{dt} = \frac{\kappa p(t)}{D + p(t)} \beta f(t) - df(t)$$

where ferret predation efficiency $\alpha$ is generalized to account for ferret maximum feeding rate, $\kappa$, and a feeding half-saturation constant, $D$.[6] A type II response function yields a positive, nonlinear relationship between the quantity of prey consumed per predator and the quantity of prey present. A type III predator response function is given by:

[1***] $$\frac{dp}{dt} = rp(t)\left(1 - \frac{p(t)}{K}\right) - \left(\frac{\kappa p(t)^2}{D^2 + p(t)^2}\right) f(t)$$

[2***] $$\frac{df}{dt} = \frac{\kappa p(t)^2}{D^2 + p(t)^2} \beta f(t) - df(t)$$

---

[6] The feeding rate is the reciprocal of the time spent feeding per prairie dog or the "handling time" of the prey.

This produces a sigmoid relationship between predation rate and prey. Figure 1a provides a visual representation of each predator response function. For robustness, we initially consider each response function as a potentially valid characterization of the relationship between ferrets and prairie dogs.

We parameterized the system using values reported in the literature (Table 1). For each system of equations, we identified and conducted a stability analysis of all the steady-state solutions (Table 2). We then numerically simulated each system of equations under a suite of initial conditions to evaluate the (pre-introduction) combinations of prairie dogs and ferrets that generate viable populations. We also measured the time it took for each population to reach $\xi$ percent from their equilibrium values. For our simulations we used an extinction threshold of 0.1 to reflect a critical threshold under which a population would not be able to recover (e.g. demographic stochasticity or Allee effects). This methodology allowed us to find the quantities of prairie dogs (pre-introduction) and introduced ferrets that yield a long-term population as well as describe the nature of the resulting steady-state equilibrium. Stability analyses were conducted in Mathematica 9.0 (Wolfram, 2015). Numerical simulations were conducted using a third order Runge-Kutta with an adaptive step size in MatLab (MatLab, 2015).

Numerical simulations implied that the type I and type II response functions are biologically intractable. For brevity, the details on these simulations are contained in Supplementary Material C. We found that it was impossible for the two species to coexist with the type II response function. The system also frequently crashed with the type I response function if there were initially too many prairie dogs, which we considered implausible. For this reason the remainder of the analysis focuses on the results from the model that uses the type III response function.

Our simulations suggest a large parameter space of initial conditions in which ferrets and prairie dogs coexist given a type III response function. The number of ferrets raised in captive breeding facilities typically falls within this range. (See, for example, Figure S4 in Supplemental Material D.) This suggests that the *ex poste* biological constraints of [3] will be satisfied for empirical applications.

*5.2. Economics*

In order to solve for the optimal time to reintroduction, $\tau^*$, we must specify cost functions for all terms in [3]. We define the costs of captive ferrets as:

$$[13] \quad c_c\left(f_c(t)\right) = l_c + a_c f_c(t) + b_c f_c(t)^{g_c}$$

The first term, $l_c$, is the lump sum cost of maintaining a captive breeding facility. The second term, $a_c$, represents costs that vary linearly with the number of ferrets, e.g. food, vaccines, and training or acclimation to wild conditions. The final terms, $b_c$ and $g_c$, are the convex costs associated with achieving a constant ferret growth rate while increasing the captive population. As the captive population grows in size, it becomes increasingly difficult for managers to expand the breeding facility and maintain a fixed growth rate, implying the costs associated with breeding are convex in nature.

After ferrets are reintroduced into the wild, there are management costs, $a_w$, associated with each ferret, namely vaccination:

[14] $$c_w(f_w(t)) = a_w f_w(t)$$

For managing wild prairie dogs, costs come in two forms. First, there are costs associated with maintaining a reintroduction site. This includes rental and maintenance of facilities, salaries and wages, vehicles, and equipment. Vaccinations and treatment for disease are often granted as single multi-year allocations and are considered part of our lump sum cost. Second, private economic costs can arise from the damages prairie dogs inflict upon the rangeland and decrease its productivity for livestock (Reading 1993; USFWS 2013). The costs of prairie dogs are therefore:

[15] $$c_p(p(t)) = l_w + b_p \frac{p(t)}{g_p + p(t)}$$

where $l_w$ is the lump sum cost of a reintroduction site and $b_p$ are the maximum damages prairie dogs impose on potential ranchland. Indirect costs saturate according to the parameter $g_p$, which is the number of prairie dogs at which 50% of ranchland productivity is lost.

Cost functions and their parameter values were calibrated from the Phoenix Zoo's black footed ferret captive breeding facility (*pers. comm.*) and the Aubrey Valley black-footed ferret reintroduction site (*pers. comm.*). We present the full list of parameter values in Table 3.

*5.3. Characterizing the analytic solution*
While it is difficult to satisfy the full suite of conditions for optimality in [6]-[12] (Cronin 2007), analyzing the necessary first order condition can provide intuition into the features that define an optimal reintroduction time. Beginning with condition [6] we highlight which combinations of parameter values may allow for an interior solution. However, when substituting realistic parameter values we find that the number of ferrets to be released at the optimal reintroduction time is a complex root. This implies that our solution for the optimal reintroduction time ($\tau^*$) is a corner.

Recall that in the analytic solution the necessary condition for the existence of an interior optimal reintroduction time, $\tau^*$, requires that the ferret management costs in captivity equal the ferret management costs in the wild, i.e. $c_c(f_c(\tau^*)) = c_w(f_w(\tau^*))$. Substituting the economic functions specified in equations [13]-[14], we derive the reintroduction time for a given set of parameters (Supplemental Material B). That is, at time $\tau^*$

[16] $$f_c^* = \frac{-(a_c - a_w) + \sqrt{(a_c - a_w)^2 - 4b_c l_c}}{2b_c}$$

where $f_c^*$ is the size of the captive ferret population ($f_c$) at the optimal reintroduction time ($\tau^*$). The solution is restricted to real, positive values. Assuming that the captive growth process is approximated with a linear term $a$ ($\frac{df_c}{dt} = a$), then a solution for the growth dynamics of captive ferrets is $f_c(t) = at + f_c^0$ where $f_c^0 = f_c(0)$ is the initial number of captive ferrets. Then the reintroduction time ($\tau^*$) is given by,

$$[17] \qquad \tau^* = \frac{f_c^* - f_c^0}{a}$$

Equations [16] and [17] imply that early reintroduction times and small translocated populations occur when the costs associated with captive breeding are high. Indeed, a necessary condition for the positivity of $f_c^*$ is that the marginal cost of the first wild ferret management exceeds the marginal cost term of the first captive-bred ferret (i.e. $a_w > a_c$). If this condition does not hold, then the optimal solution is to reintroduce captive ferrets immediately - per ferret, it is cheaper to let ferrets grow in the wild than in captivity (naturally, this assumes that the extinction threat has been addressed). We find that the lump sum cost of captive ferret management ($l_c$) and the quadratic coefficient for convex ferret management costs in the wild ($b_c$) are negatively correlated with $f_c^*$ and $\tau^*$. The more expensive it is to raise ferrets in captivity, the sooner the optimal release time and population size. Substituting realistic parameter values into equations [16] and [17] yields a complex root, suggesting that the optimal reintroduction time is a corner solution. This result is confirmed by numerical analysis below (Figures 2 and 3).

*5.4. Numerical analysis*
In order to evaluate the intuition gained from analyzing the necessary first order condition and overcome some of the limitations of the analytic solution, we construct a numerical scheme to calculate the optimal reintroduction time using brute-force, simulation methods. The numerical scheme allows us to provide an accurate solution under different modeling scenarios.
 For a given set of initial conditions (captive ferrets and wild prairie dogs) and candidate reintroduction times, the numerical scheme:

1. Solves the population equations up to the candidate reintroduction time
   - Calculates management costs from the initial time until the reintroduction time
   - Stores terminal values of captive ferret and wild prairie dogs
2. Solves the populations equations from the reintroduction time to the terminal time
   - Calculates the management costs over this cost horizon
3. Calculate the total costs of the management program (pre- and post-introduction)

In general we find that it is optimal to reintroduce ferrets early in the management program. Figures 2 and 3 show that reintroduction should occur immediately, regardless of changes in the cost of managing captive or wild ferrets, maximum prairie dog damages, and the captive ferret growth rate. Only when prairie dogs provide positive benefits (negative damages) is the candidate reintroduction time definitively interior (Figure 3b). These results are in agreement with our analytic analysis above.

Holding other economic parameters constant, it seems that the reintroduction time is relatively insensitive to changes in captive ferret growth rates. Additional costs associated with breeding ferrets at a faster rate in captivity are outweighed by the damages caused by prairie dogs in the wild.

Figures 2 and 3 also illustrate intervals of candidate reintroduction times in which managers can reintroduce ferrets and not incur significant costs. This result implies the existence of reintroduction thresholds which, if crossed (e.g. ferrets are reintroduced too late) lead to cost penalties. For instance, when captive growth rates vary between 0 and 5 the cost of the total management program changes insignificantly if managers reintroduce ferrets within 50 years (Figure 2a). But waiting until 100 years results in substantially larger costs.

## 6. Discussion

Our analyses indicate that reintroduction should occur as soon as possible, i.e. a corner solution (Figures 2, 3). While we initially derived a general set of analytic conditions for an optimal reintroduction time, our results suggest an interior solution cannot be optimal. This important result implies that there are only two courses of action to protect ferrets: reintroduce immediately or maintain the captive population indefinitely. The sensitivity analyses illustrated in Figures 2 and 3 provide robust evidence that reintroduction should begin immediately.

To this end, our results indicate that we should direct our conservation efforts to address *in situ* problems for reintroduction, namely threats to ferret survival in the wild. Ferrets are sensitive animals and their populations experience sharp responses to changes in their prey base, environmental conditions, and disease (USFWS 2013). Of particular concern are canine distemper and sylvatic plague, diseases spread through animal-to-animal contact and indirectly via flea vectors or contaminated material (USFWS 2013; Antolin et al. 2002). Current efforts are underway to increase ferret survival including acclimatization, training, and the development of vaccines and practices to manage disease (USFWS 2013; Abbott et al. 2012).

In focusing on the bioeconomics of translocation, this analysis touches on the issue of *ex situ* management. The major theme in bioeconomics has been the optimal extraction or *in situ* management of natural resources (Tschirhart 2009; Conrad and Clark 1987; Conrad 1999). In contrast, translocation is an *ex situ* control that must account for *ex ante* and *ex post* population dynamics to determine its success. Models in the existing literature largely assume direct control over the target species, e.g. harvest or stocking.[7] Furthermore, traditional solution methods in optimal control theory were developed for resource management problems involving the dynamic allocation of resources, i.e. how investments in *in situ* resource management should change over time (Clark 1976). Resource problems involving one-off investments do not fit into this framework.  For this type of resource economics problem, the control variable is effectively a single-use, impulse control.[8]

---

[7] Reintroduction programs for endangered species often depend on captive breeding programs that risk creating genetic bottlenecks. Thus, much of the literature on the biology of managing captive populations focuses on genetic variability (Seal et al. 1989).

Many model extensions are beyond the scope of this paper and are left for future work. We discuss three of them here.  First, our model provides an idealized representation of the ferret and prairie dog system in that it is deterministic and does not capture fluctuations in predator or prey populations due to seasonal variability in temperature or rainfall.  In reality, small fluctuations in prairie dog biomass can have large effects on ferret populations and ferrets themselves are sensitive to environmental conditions (USFWS 2013).  Second, our models are not spatial.  Both species disperse from their natal colonies to form additional territories (USFWS 2013), which has implications for the conservation of each species.  Metacommunity theory suggests that species dispersal can replace locally extinct species populations or prevent local extinction by replenishing declining populations (Brown and Kodric-Brown 1977; Shmida and Wilson 1985; Holt 1985; Pulliam 1988).  The sensitive nature of ferret populations suggests that a network of spatially-connected populations would increase the resilience of the species to spatial-temporal environmental shocks.  Finally, disease is an on-going management problem, with insecticides and vaccines currently being used to protect reintroduced populations.  In our model disease dynamics can be implemented in a straightforward way, though it will change the necessary first order conditions for optimization.  One approach is to treat disease states as separate species classes or compartments, e.g. the classic Susceptible-Infected-Recovered (SIR) epidemiological model (Lenhart and Workman 2007).

## 7. Conclusion

Generally we find that it is optimal to reintroduce ferrets early on in the reintroduction program, particularly if the costs of prairie dogs and thus the benefits of ferret ecosystem services are high. Based on the set of parameters examined, it looks as though the optimal duration of the captive breeding program is short relative to current ferret management practices.  We attribute part of this difference to on-going habitat threats, rather than purely inefficiencies in the ferret translocation strategy. However, our results imply that reintroduction should be delayed only until *in situ* threats to introduction are addressed.

More generally, our research suggests that captive breeding programs with expensive marginal animal costs should be short-lived with the animals quickly reintroduced into the wild. An example of an efficient captive breeding program may be seen in the case of the California Island Fox, whose wild population was effectively extirpated by invasive golden eagles. After the foxes were placed on the U.S. endangered species list in 2004, captive breeding populations were set up. After the eagles were removed, the captive-bred population was quickly reintroduced to the wild and has since recovered sufficiently to be removed from the endangered species list.

---

[8] See Dixit and Pindyck (1994) and Conrad (1999) for examples in the optimal stopping time of market investments or the optimal rotation time of timber harvest.  Like timber harvesting, translocation is an impulse control that can be used only once (or at least at discrete individuals).

## 8. Acknowledgements

We would like to thank the Arizona Fish and Game Department for cost and population census data of Arizona's Aubrey Valley reintroduction site. We would also like to acknowledge the Phoenix Zoo for providing cost and population data for the Phoenix Zoo captive breeding program; calibration of our models would not have been possible without their support. KRS was supported by the NSF Alliance for Building Faculty Diversity Postdoctoral Fellowship [NSF 733 grant number DMS-0946431].

Table 1. Summary of biological parameters.

| Parameter | Value | Interpretation | Units | Reference |
|---|---|---|---|---|
| $a$ | 0.0725 | captive ferret growth rate | time$^{-1}$ | Phoenix Zoo captive ferret breeding program (Supplementary Material D) |
| $r$ | 0.6 | prairie dog reproduction rate | time$^{-1}$ | Hoogland et al. (1987); Klebanoff et al. (1991) |
| $K$ | 52650 | prairie dog carrying capacity | p biomass | Holmes (2008) |
| $\alpha$ | 0.006 | ferret feeding efficiency | $\dfrac{1}{\text{f biomass} * \text{time}}$ | Biggins et al. (1993) |
| $\beta$ | 0.032 | prairie dog to ferret biomass conversion efficiency | $\dfrac{\text{f biomass}}{\text{p biomass}}$ | Biggins et al. (1993); Wilson and Ruff (1999) |
| $d$ | 0.6 | ferret mortality rate | time$^{-1}$ | Klebanoff et al. (1991) |
| $h$ | 0.0092 | ferret handling time | $\dfrac{\text{f biomass} * \text{time}}{\text{p biomass}}$ | Biggins et al. (1993) |
| $\kappa$ | $1/h$ | ferret maximum feeding rate | $\dfrac{\text{p biomass}}{\text{f biomass} * \text{time}}$ | - |
| $D$ | $1/\alpha h$ | feeding half-saturation constant | p biomass | - |

Prairie dog carrying capacity and ferret feeding efficiency were calibrated to the Wolf Creek Restoration Site, Colorado (Holmes 2008). Carrying capacity was set as the maximum recorded population of prairie dogs on the 7,700 hectare property. Feeding efficiency was approximated as the annual average number of prairie dogs consumed for a single ferret (109) (Biggins et al. 1993) divided by the initial population of prairie dogs at the Wolf Creek site (18,000) (Holmes 2008). Prairie dog to ferret biomass conversion efficiency was found by the ratio of the average ferret litter size (3.5) (Wilson and Ruff 1999) per average number of prairie dogs consumed (109) (Biggins et al. 1993). Handling time was calculated as the average number of prairie dogs that are consumed annually by a single ferret (1/109).

Table 2. Biological steady states and stability of equilibria.

| | Equilibria | Restrictions | Eigenvalues | Stability of Equilibria |
|---|---|---|---|---|
| **Type I** | $p_1^* = 0, f_1^* = 0$ | - | -0.6, 0.6 | Saddle |
| | $p_2^* = K, f_2^* = 0$ | - | 9.544, -0.6 | Saddle |
| | $p_3^* = \dfrac{d}{\alpha\beta}$ $f_3^* = \dfrac{r}{\alpha}\left(1 - \dfrac{p_3^*}{K}\right)$ | $p_3^* < K$ | -0.0177 + 0.582i $\\$ -0.1777 + 0.582i | Stable spiral |
| **Type II** | $p_1^* = 0, f_1^* = 0$ | - | -0.6, 0.6 | Saddle |
| | $p_2^* = K, f_2^* = 0$ | - | 2.002, -0.6 | Saddle |
| | $p_3^* = \dfrac{dD}{\kappa\beta - d}$ $f_3^* = \dfrac{r}{\kappa}\left(1 - \dfrac{p_3^*}{K}\right)(D + p_3^*)$ | $\kappa\beta > d$ $p_3^* < K$ | 0.0263 + 0.526i $\\$ 0.0263 - 0.526i | Unstable spiral |
| **Type III** | $p_1^* = 0, f_1^* = 0$ | - | -0.6, 0.6 | Saddle |
| | $p_2^* = K, f_2^* = 0$ | - | 2.528, -0.6 | Saddle |
| | $p_3^* = D\dfrac{\sqrt{d}}{\sqrt{\kappa\beta - d}}$ $f_3^* = r\left(1 - \dfrac{p_3^*}{K}\right)\dfrac{D^2 + (p_3^*)^2}{\kappa p_3^*}$ | $\kappa\beta > d$ $p_3^* < K$ | -0.213 + 0.676i $\\$ -0.213 - 0.676i | Stable spiral |

Parameter values in Table 1 were used to evaluate the stability of equilibria. For each predator response function the number of ferrets and prairie dogs at the interior steady state were: 94 ferrets, 3114 prairie dogs (type I); 112 ferrets, 3758 prairie dogs (type II); and 223 ferrets, 8263 prairie dogs (type III).

Table 3. Summary of cost parameters.

| Parameter | Value | Interpretation |
| --- | --- | --- |
| $l_c$ | $185,000 | lump sum cost of a maintaining a captive breeding facility |
| $l_w$ | $200,000 | lump sum costs of maintaining a reintroduction site |
| $a_c$ | $400 | per ferret cost in captivity |
| $a_w$ | variable 45 [50, 750] | per ferret cost in the wild |
| $b_c$ | variable 1 [0, 750] | scalar of convex costs of ferrets in captivity |
| $b_p$ | variable $574,560 [0, 750000] | maximum damages of prairie dogs to surrounding pasture or rangeland |
| $g_c$ | 2 | degree of convexity in convex costs of ferrets in captivity |
| $g_p$ | 0.75*K | number of prairie dogs where half maximum damage occurs |

Parameter values were calibrated from data from the Phoenix Zoo's black-footed ferret captive breeding facility and the Aubrey Valley, Arizona reintroduction site. Costs on an annual basis. The damages of prairie dogs were calculated as the average value per acre of pastureland in 2015 ($1340) (USDA 2015) multiplied by the average farm size (includes rangeland and pasture) from 2000 to 2010 (432) (USDA 2011). Brackets indicate the range of values tested in sensitivity analyses.

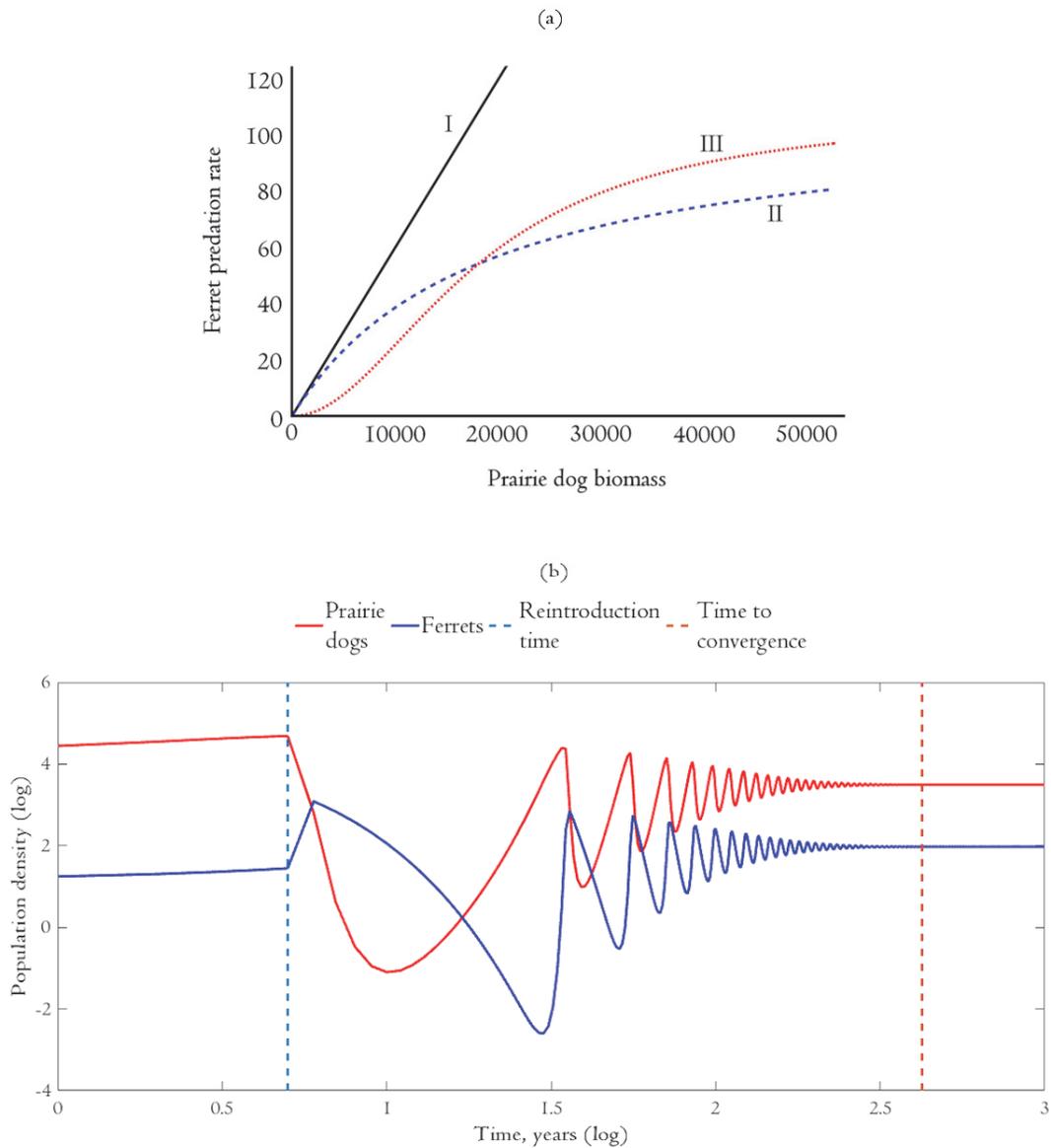

Figure 1. Ferret predator response functions (a) and example population dynamics for a reintroduction scheme (b). (a) Color and style indicate predator response functions: type I (black, solid), type II (blue, dashed), and type III (red, dotted). (b) Color indicates prairie dog (red) and ferret (blue) population levels. The dashed lines indicate time points of reintroduction (blue) and equilibrium (red) respectively.

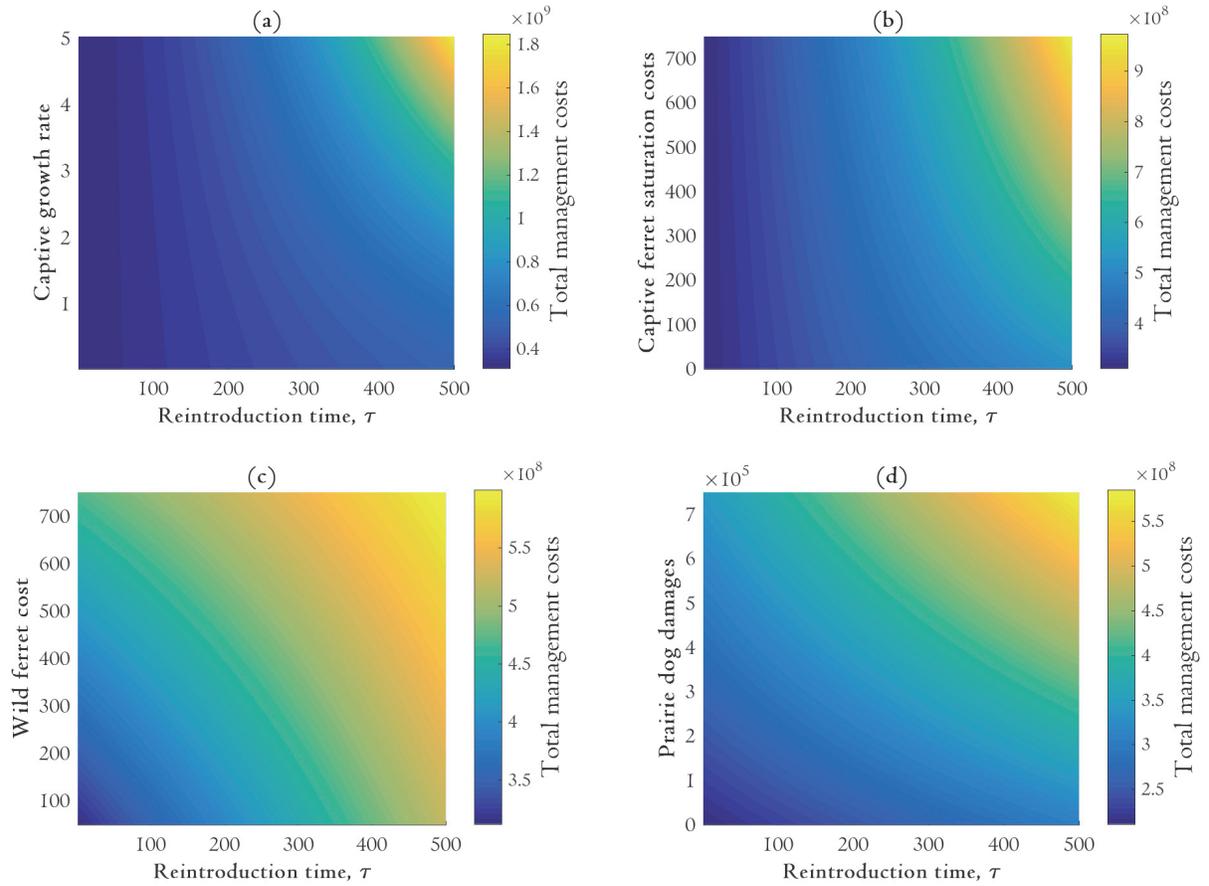

Figure 2. Sensitivity of optimal reintroduction time to model parameters: (a) captive ferret growth rate, (b) captive ferret saturation costs, (c) wild ferret costs, and (d) maximum prairie dog damages. Color indicates the total cost of the management program. Note the difference in scales of total cost. Results are for the numerical algorithm.

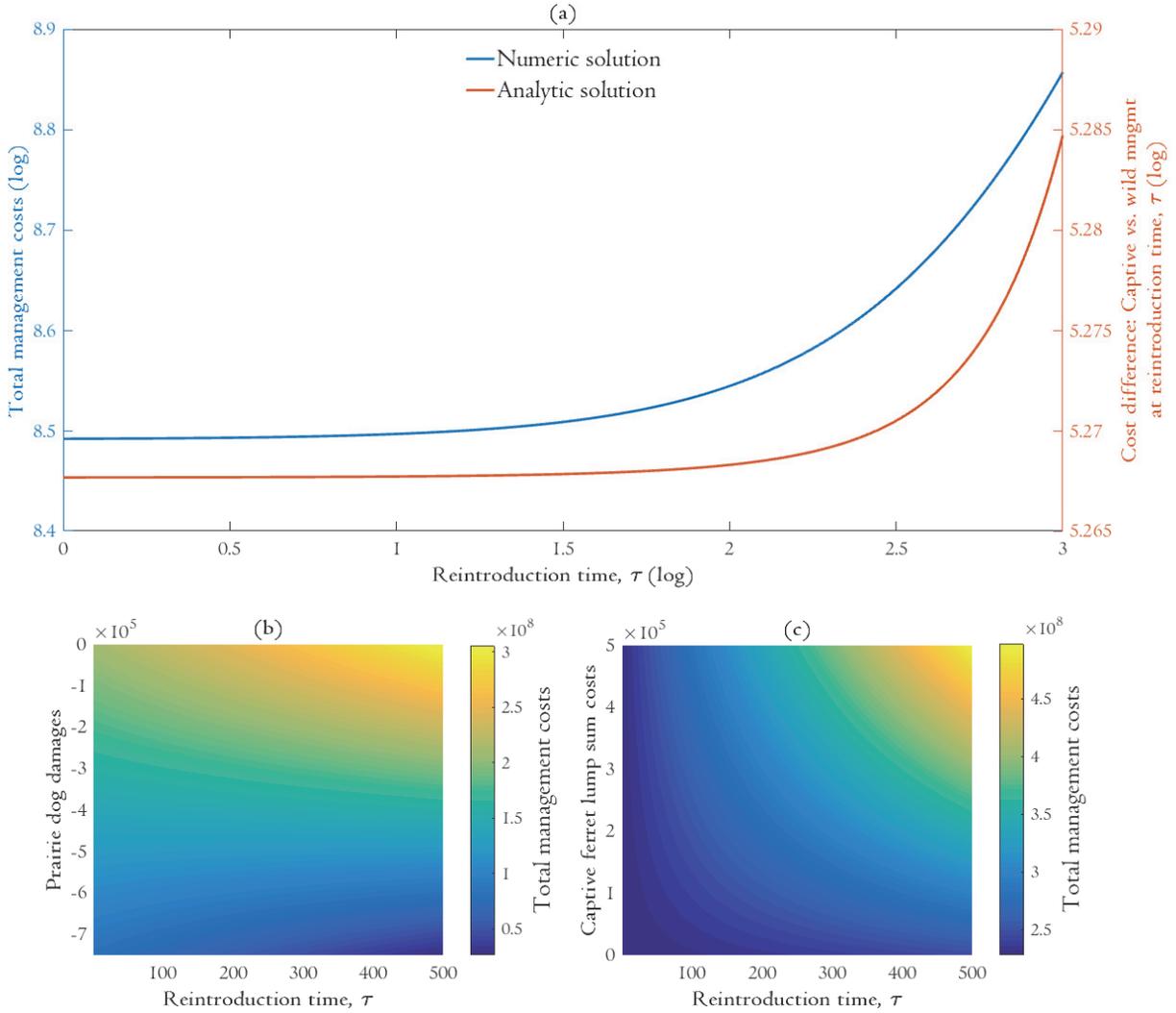

Figure 3. Comparison of numeric and analytic methods (a) and sensitivity of the optimal reintroduction time to maximum prairie dog benefits (negative damages) (b) and captive ferret lump sum costs (c).[9] In (a), color indicates the method: analytical solution (red) and numeric solution (blue). In the analytic scheme, the optimal reintroduction time is the candidate reintroduction time in which the costs of raising ferrets in captivity are the same as keeping ferrets in the wild. In the numeric scheme, this is the time that minimizes the total costs of the management program. In (b-c), decreasing the net costs of pre-reintroduction leads to a later optimal reintroduction time. Note that maximum prairie dog damages in (b) are negative (e.g. benefits).

---

[9] The parameters in (a, b) are our benchmark parameters. In (c) the following parameter values deviate from the benchmark: $a_c = 200$ and $b_p = 100,000$.

**Supplementary Material A. Analytical derivation of reintroduction time**

We adopt a first-principles approach in deriving the optimal reintroduction time. Often we take the necessary first-order condition(s), adjoint equations, and transversality conditions as given when in fact they are derived by evaluating the change in the Lagrangian or Hamiltonian with respect to a small change in the control (Clark 2010; Conrad and Clark 1987; Lenhart and Workman 2007).

In this section, we first present the first-principles approach to optimal control theory following the classic example of Conrad and Clark (1987). We then apply the method to derive the optimal reintroduction time of ferrets.

*A1. A classic example*

Following Conrad and Clark (1987), let $y(t)$ and $x(t)$ denote the control and state variables at time $t$. Define the value function describing the costs and benefits of the state and control variables as $V(x(t), y(t), t)$. Then the Lagrangian expression takes the form,

[A1] $$H(x(t), y(t), \lambda(t), t) = V(x(t), y(t), t) + \lambda(t) g(x(t), y(t))$$

[A2] $$L = \int_o^T \left[ H(x(t), y(t), \lambda(t), t) + \frac{d\lambda}{dt} x(t) \right] dt + \varphi(x(T)) - \lambda(T) x(T) + \lambda(0) x(0)$$

where $H(x(t), y(t), \lambda(t), t)$ is a continuous-time Hamiltonian, $\lambda(t)$ is the costate variable, $g(x(t), y(t))$ is the equation of motion for the state variable, $\varphi(x(T))$ is the scrap value function, and $T$ indicates the terminal time.

Conrad and Clark (1987), Lenhart and Workman (2007), and Clark (2010) note that a change in the control causes a subsequent perturbation in the state variable such that:

[A3] $$y(t) \to y^\varepsilon(t) = y(t) + \varepsilon h(t)$$

[A4] $$x(t) \to x^\varepsilon(t)$$

where $h(t)$ is a piecewise continuous variation function and $\varepsilon$ is an arbitrarily small constant. Thus $y^\varepsilon(t)$ reflects a small change in the control and $x^\varepsilon(t)$ is the new state in response to the change in the control variable. We then derive the perturbed ($L_\varepsilon$), and evaluate a minimum/maximum by equating the difference between $L$ and $L_\varepsilon$ to zero. Thus,

[A5] $$\Delta L = \lim_{\varepsilon \to 0} \frac{L_\varepsilon - L}{\varepsilon} = \left. \frac{\partial L_\varepsilon}{\partial \varepsilon} \right|_{\varepsilon=0} = 0$$

Equation [A5] encompasses the direct change that the control variable has on the Lagrangian, as well as the secondary changes in the state and costate variables due to the change in the control. In order to ensure optimality, [A5] must hold for every control applied to the system. From [A5] one may derive the general set of equations normally referred to as the necessary first order condition (e.g. the maximum principle) and the adjoint and transversality conditions. See Conrad and Clark (1987) (pgs. 25-31) for details.

*A2. Deriving the optimal reintroduction time*

Following a first-principles approach, we construct our Lagrangian, compute the corresponding partial derivative, and solve for the conditions for optimality. While the general approach is the same, there are some distinct differences between the classic problem and our application. Most notable is the control variable. In contrast to harvest, our control variable is time which does not have a direct effect on the state variables. Further, we make our decision only once during the time horizon such that, at the moment of reintroduction, the marginal benefits of reintroduction equal the marginal costs of delaying it. The associated parallel in traditional optimal control theory is to choose harvest at each moment in time such that the marginal benefits of harvest equal the marginal costs (Conrad and Clark 1987; Conrad 1999; Clark 2010).

*A2.1. Constructing L and $L_\varepsilon$*

As a reminder, our objective function is:

[A6]
$$\min_\tau \int_0^\tau \left[ c_c(f_c(t)) + c_p(p_c(t)) \right] e^{-\delta t} dt + \int_\tau^T \left[ c_w(f_w(t)) + c_p(p_w(t)) \right] e^{-\delta t} dt$$
$$+ \varphi(T, f_w(T), p_w(T))$$

subject to:
$$\frac{df_c}{dt}, \frac{df_w}{dt}, \frac{dp_c}{dt}, \frac{dp_w}{dt}$$
$$f_c(0), p_c(0)$$
$$f_c(\tau) = f_w(\tau)$$
$$p_c(\tau) = p_w(\tau)$$

where the stock of ferrets in captivity is denoted by $f_c(t)$, ferrets in the wild by $f_w(t)$, and prairie dogs in the wild pre- and post- reintroduction by $p_c(t)$ and $p_w(t)$. Candidate reintroduction and the terminal times are given by $\tau$ and $T$ respectively. Costs of ferrets in captivity/wild and prairie dogs in the wild are denoted by $c_c(f_c(t))$, $c_w(f_w(t))$, $c_p(p_c(t))$, and $c_p(p_w(t))$ respectively. The scrap value and the discount rate are indicated by $\varphi$ and $\delta$. We further evaluate *ex poste* the minimum and maximum number of ferrets and prairie dogs present at reintroduction that yield a long-term equilibrium of both species in the wild ( $f_{min} \leq f_c(\tau) \leq f_{max}$ and $p_{min} \leq p_c(\tau) \leq p_{max}$). These is treated as a biological constraint to the problem. Solving [A6] subject to the biological constraints of the system defines the conditions to solve for an optimal reintroduction time $\tau^*$.

For simplicity of notation, let

$$\frac{df_c}{dt} = g_1(f_c(t))$$

$$\frac{dp_c}{dt} = g_2(p_c(t))$$

$$\frac{df_w}{dt} = g_3(p_w(t), f_w(t))$$

$$\frac{dp_w}{dt} = g_4(p_w(t), f_w(t))$$

$$H_c(t, f_c(t), p_c(t)) = \left[c_c(f_c(t)) + c_p(p_c(t))\right]e^{-\delta t} + \lambda_1 g_1 + \mu_1 g_2$$

$$H_w(t, f_w(t), p_w(t)) = \left[c_w(f_w(t)) + c_p(p_w(t))\right]e^{-\delta t} + \lambda_2 g_3 + \mu_2 g_4$$

where the value of an extra unit of ferrets (prairie dogs) pre- and post-reintroduction are given by $\lambda_1$ and $\lambda_2$ ($\mu_1$ and $\mu_2$) respectively.

Then the present-value Lagrangian describing the total cost of managing ferrets and prairie dogs in perpetuity simplifies to:

[A7] $$L = \int_0^\tau \left[H_c - \lambda_1 \frac{df_c}{dt} - \mu_2 \frac{dp_c}{dt}\right]dt + \int_\tau^T \left[H_w - \lambda_2 \frac{df_w}{dt} - \mu_2 \frac{dp_w}{dt}\right]dt + \varphi$$

Note that we have dropped the input arguments in the functions above for simpler notation. Following Conrad and Clark (1987) and Clark (2010), integration by parts allows us to rewrite the Lagrangian as,

[A8] $$L = \begin{aligned}&\int_0^\tau \left[H_c + \frac{d\lambda_1}{dt}f_c + \frac{d\mu_1}{dt}p_c\right]dt + \int_\tau^T \left[H_w + \frac{d\lambda_2}{dt}f_w + \frac{d\mu_2}{dt}p_w\right]dt + \varphi \\ &-\left[\lambda_1(\tau)f_c(\tau) - \lambda_1(0)f_c(0)\right] - \left[\mu_1(\tau)p_c(\tau) - \mu_1(0)p_c(0)\right] \\ &-\left[\lambda_2(T)f_w(T) - \lambda_2(\tau)f_w(\tau)\right] - \left[\mu_2(T)p_w(T) - \mu_2(\tau)p_w(\tau)\right]\end{aligned}$$

The reintroduction time for the Lagrangian in equation [A8] is $\tau$. Evaluating the Lagrangian for a small change $\varepsilon$ in the control variable, we obtain

$$\int_0^{\tau+\varepsilon}\left[H_c+\frac{d\lambda_1}{dt}f_c+\frac{d\mu_1}{dt}p_c\right]dt+\int_{\tau+\varepsilon}^T\left[H_w^\varepsilon+\frac{d\lambda_2}{dt}f_w^\varepsilon+\frac{d\mu_2}{dt}p_w^\varepsilon\right]dt+\varphi^\varepsilon$$

[A9] $\quad L_\varepsilon = \quad -\left[\lambda_1(\tau+\varepsilon)f_c(\tau+\varepsilon)-\lambda_1(0)f_c(0)\right]-\left[\mu_1(\tau+\varepsilon)p_c(\tau+\varepsilon)-\mu_1(0)p_c(0)\right]$

$\qquad\qquad -\left[\lambda_2(T)f_w^\varepsilon(T)-\lambda_2(\tau+\varepsilon)f_w^\varepsilon(\tau+\varepsilon)\right]-\left[\mu_2(T)p_w^\varepsilon(T)-\mu_2(\tau+\varepsilon)p_w^\varepsilon(\tau+\varepsilon)\right]$

where the new reintroduction time is $\tau+\varepsilon$. We have made the following transformations to indicate the changes due to the reintroduction time: $\tau \to \tau+\varepsilon$, $f_w(t) \to f_w^\varepsilon(t)$, $p_w(t) \to p_w^\varepsilon(t)$, $H_w^\varepsilon = H_w\left(t, f_w^\varepsilon(t), p_w^\varepsilon(t), \lambda_3(t), \lambda_4(t)\right)$ and $\varphi^\varepsilon = \varphi\left(f_w^\varepsilon(T), p_w^\varepsilon(T)\right)$. There exist no $H_c^\varepsilon$ because the pre-reintroduction trajectories are not affected by a change in $\tau$. The solution only uses information about the system when a control is applied to the system, e.g. when ferrets are reintroduced into the wild (Clark 2010; Conrad 1999; Conrad and Clark 1987; Lenhart and Workman 2007).[1]

## A2.2. Deriving the change in the Lagrangian ($\Delta L$)

Following Conrad and Clark (1987), Lenhart and Workman (2007), and Clark (2010), at an optimal reintroduction time the change in the Lagrangian for a small change in the control must go to zero. This is equivalent to evaluating the partial derivative of [A9] with respect to the control at zero (Lenhart and Workman 2007). That is, at the optimal reintroduction time the following hold:

[A10] $$\Delta L = \lim_{\varepsilon\to 0}\frac{L_\varepsilon - L}{\varepsilon} = \left.\frac{\partial L_\varepsilon}{\partial \varepsilon}\right|_{\varepsilon=0} = 0$$

---

[1] It is important to identify the effect that our control variable, $\tau$, has on the state variables that make up the ecological model. In order to optimally choose the reintroduction time we must understand how a change in the reintroduction time alters not only the abundances of ferrets and prairie dogs, but also all possible changes in the population trajectories of each species. Figure S1 illustrates how our state variables change in response to a small change in the reintroduction time. As we perturb the control from $\tau$ to $\tau+\varepsilon$, the following hold:

1. Pre-reintroduction. Ferrets exhibit constant growth and prairie dogs, in the absence of predation pressure, exhibit logistic growth. While the bounds have changed (and by extension the final value of the state variables), each species remains on the same trajectory as before (Cronin, 2007).

2. Post-reintroduction. Each of the state variables are perturbed, but because the predator-prey system dynamics are asymptotically stabilizing (an important assumption) the change in the orbits are minimal (Cronin, 2007). As $\tau$ shifts to $\tau+\varepsilon$, the states shift from $f_w(t)$ to $f_w^\varepsilon(t)$ and $p_w(t)$ to $p_w^\varepsilon(t)$ respectively.

Therefore, although the abundances of ferrets and prairie dogs change with the reintroduction time, the population trajectories do not.

Explicitly indicating the effect of changing the reintroduction time yields equation [5] in the main text. That is,

[A11]
$$\Delta L = \frac{\partial L}{\partial \tau} + \left[\frac{\partial L}{\partial f_c} + \frac{\partial L}{\partial p_c}\right] + \left[\frac{\partial L}{\partial f_w} + \frac{\partial L}{\partial p_w}\right] = 0$$

where the first term in [A11] is the direct change in the Lagrangian for a small change in the reintroduction time, and the bracketed terms reflect the secondary effects of changing the reintroduction time on wild ferret and prairie dog abundances (and subsequently the Lagrangian).

By substituting $L$ and $L_\varepsilon$ into [A10] and taking the limit as $\varepsilon \to 0$, we may write the partial derivative of $L_\varepsilon$ as,

[A12]
$$\left.\frac{\partial L_\varepsilon}{\partial \varepsilon}\right|_{\varepsilon=0} = \left\{ \begin{array}{l} \left[H_c + \frac{d\lambda_1}{dt}f_c + \frac{d\mu_1}{dt}p_c\right]_{t=\tau+\varepsilon} - \left[\frac{d\lambda_1}{dt}f_c + \lambda_1 \frac{df_c}{dt}\right]_{t=\tau+\varepsilon} - \left[\frac{d\mu_1}{dt}p_c + \mu_1 \frac{dp_c}{dt}\right]_{t=\tau+\varepsilon} \\ + \int_{\tau+\varepsilon}^{T}\left[\frac{\partial H_w^\varepsilon}{\partial f_w}\frac{\partial f_w^\varepsilon}{\partial \varepsilon} + \frac{\partial H_w^\varepsilon}{\partial p_w}\frac{\partial p_w^\varepsilon}{\partial \varepsilon} \frac{d\lambda_2}{dt}\frac{\partial f_w^\varepsilon}{\partial \varepsilon} + \frac{d\mu_2}{dt}\frac{\partial p_w^\varepsilon}{\partial \varepsilon} +\right]dt - \left[H_w^\varepsilon + \frac{d\lambda_2}{dt}f_w^\varepsilon + \frac{d\mu_2}{dt}p_w^\varepsilon\right]_{t=\tau+\varepsilon} \\ -\left[\lambda_2 \frac{\partial f_w^\varepsilon}{\partial \varepsilon}\right]_{t=T} + \left[\frac{d\lambda_2}{dt}f_w^\varepsilon + \lambda_2\left(\frac{\partial f_w^\varepsilon}{\partial \varepsilon} + \frac{df_w^\varepsilon}{dt}\right)\right]_{t=\tau+\varepsilon} \\ -\left[\mu_2 \frac{\partial p_w^\varepsilon}{\partial \varepsilon}\right]_{t=T} + \left[\frac{d\mu_2}{dt}p_w^\varepsilon + \mu_2\left(\frac{\partial p_w^\varepsilon}{\partial \varepsilon} + \frac{dp_w^\varepsilon}{dt}\right)\right]_{t=\tau+\varepsilon} \\ + \left[\frac{\partial \varphi^\varepsilon}{\partial f_w}\frac{\partial f_w^\varepsilon}{\partial \varepsilon} + \frac{\partial \varphi^\varepsilon}{\partial p_w}\frac{\partial p_w^\varepsilon}{\partial \varepsilon}\right]_{t=T} \end{array} \right\}_{\varepsilon=0}$$

where we have made the following substitutions in [A12]:

(by Leibniz integral rule)
$$\frac{\partial}{\partial \varepsilon}\left\{\int_0^{\tau+\varepsilon}\left[H_c + \frac{d\lambda_1}{dt}f_c + \frac{d\mu_1}{dt}p_c\right]dt\right\} = \left[H_c + \frac{d\lambda_1}{dt}f_c + \frac{d\mu_1}{dt}p_c\right]_{t=\tau+\varepsilon}$$

(by Leibniz integral rule)

$$\frac{\partial}{\partial \varepsilon}\left\{\int_{\tau+\varepsilon}^{T}\left[H_w^\varepsilon + \frac{d\lambda_2}{dt}f_w^\varepsilon + \frac{d\mu_2}{dt}p_w^\varepsilon\right]dt\right\}$$

$$= \int_{\tau+\varepsilon}^{T}\left[\frac{\partial H_w^\varepsilon}{\partial \varepsilon} + \frac{d\lambda_2}{dt}\frac{\partial f_w^\varepsilon}{\partial \varepsilon} + \frac{d\mu_2}{dt}\frac{\partial p_w^\varepsilon}{\partial \varepsilon}\right]dt - \left[H_w^\varepsilon + \frac{d\lambda_2}{dt}f_w^\varepsilon + \frac{d\mu_2}{dt}p_w^\varepsilon\right]\bigg|_{t=\tau+\varepsilon}$$

$$= \int_{\tau+\varepsilon}^{T}\left[\frac{\partial H_w^\varepsilon}{\partial f_w}\frac{\partial f_w^\varepsilon}{\partial \varepsilon} + \frac{\partial H_w^\varepsilon}{\partial p_w}\frac{\partial p_w^\varepsilon}{\partial \varepsilon} + \frac{d\lambda_2}{dt}\frac{\partial f_w^\varepsilon}{\partial \varepsilon} + \frac{d\mu_2}{dt}\frac{\partial p_w^\varepsilon}{\partial \varepsilon}\right]dt - \left[H_w^\varepsilon + \frac{d\lambda_2}{dt}f_w^\varepsilon + \frac{d\mu_2}{dt}p_w^\varepsilon\right]\bigg|_{t=\tau+\varepsilon}$$

(by the product rule and chain rule)

$$\frac{\partial}{\partial \varepsilon}\{\mu_1(\tau+\varepsilon)p_c(\tau+\varepsilon)\} = \frac{d\mu_1}{dt}\bigg|_{t=\tau+\varepsilon}p_c(\tau+\varepsilon) + \mu_1(\tau+\varepsilon)\frac{dp_c}{dt}\bigg|_{t=\tau+\varepsilon}$$

$$= \left[\frac{d\mu_1}{dt}p_c + \mu_1\frac{dp_c}{dt}\right]\bigg|_{t=\tau+\varepsilon}$$

$$\frac{\partial}{\partial \varepsilon}\{\mu_2(T)p_w^\varepsilon(T)\} = \mu_2(T)\frac{\partial p_w^\varepsilon}{\partial \varepsilon}\bigg|_{t=T}$$

$$= \left[\mu_2\frac{\partial p_w^\varepsilon}{\partial \varepsilon}\right]\bigg|_{t=T}$$

(by product rule, chain rule, and definition of a total derivative)

$$\frac{\partial}{\partial \varepsilon}\{\mu_2(\tau+\varepsilon)p_w^\varepsilon(\tau+\varepsilon)\} = \frac{d\mu_2}{dt}\bigg|_{t=\tau+\varepsilon}p_w^\varepsilon(\tau+\varepsilon) + \mu_2(\tau+\varepsilon)\left[\frac{\partial p_w^\varepsilon}{\partial \varepsilon}\bigg|_{t=\tau+\varepsilon} + \frac{dp_w^\varepsilon}{dt}\bigg|_{t=\tau+\varepsilon}\right]$$

$$= \left[\frac{d\mu_2}{dt}p_w^\varepsilon + \mu_2\left(\frac{\partial p_w^\varepsilon}{\partial \varepsilon} + \frac{dp_w^\varepsilon}{dt}\right)\right]\bigg|_{t=\tau+\varepsilon}$$

(by the definition of a total derivative)

$$\frac{\partial}{\partial \varepsilon}\{\varphi^\varepsilon\} = \left[\frac{\partial \varphi^\varepsilon}{\partial f_w}\frac{\partial f_w^\varepsilon}{\partial \varepsilon} + \frac{\partial \varphi^\varepsilon}{\partial p_w}\frac{\partial p_w^\varepsilon}{\partial \varepsilon}\right]\bigg|_{t=T}$$

## A2.3. Simplifying the change in the Lagrangian ($\Delta L$)

In order to simplify the partial derivative of $L_\varepsilon$ (equation [A12]), note the following simplifications:

(by definition of $H_c$)

$$\left[H_c + \frac{d\lambda_1}{dt} f_c + \frac{d\mu_1}{dt} p_c\right]_{t=\tau+\varepsilon} - \left[\frac{d\lambda_1}{dt} f_c + \lambda_1 \frac{df_c}{dt}\right]_{t=\tau+\varepsilon} - \left[\frac{d\mu_1}{dt} p_c + \mu_1 \frac{dp_c}{dt}\right]_{t=\tau+\varepsilon}$$
$$= \left[\left[c_c(f_c) + c_p(p_c)\right]e^{-\delta t}\right]_{t=\tau+\varepsilon}$$

(by definition of $H_w$)

$$\left[H_w^\varepsilon + \frac{d\lambda_2}{dt} f_w^\varepsilon + \frac{d\mu_2}{dt} p_w^\varepsilon\right]_{t=\tau+\varepsilon} - \left[\frac{d\lambda_2}{dt} f_w^\varepsilon + \lambda_2 \frac{df_w^\varepsilon}{dt}\right]_{t=\tau+\varepsilon} - \left[\frac{d\mu_2}{dt} p_w^\varepsilon + \mu_2 \frac{dp_w^\varepsilon}{dt}\right]_{t=\tau+\varepsilon}$$
$$= \left[\left[c_w(f_w^\varepsilon) + c_p(p_w^\varepsilon)\right]e^{-\delta t}\right]_{t=\tau+\varepsilon}$$

Thus, we reduce the partial derivative of $L_\varepsilon$ (equation [A12]) to,

[A13]
$$\left.\frac{\partial L_\varepsilon}{\partial \varepsilon}\right|_{\varepsilon=0} = \left\{ \begin{aligned} &\left[\left[c_c(f_c) + c_p(p_c)\right]e^{-\delta t}\right]_{t=\tau+\varepsilon} - \left[\left[c_w(f_w^\varepsilon) + c_p(p_w^\varepsilon)\right]e^{-\delta t}\right]_{t=\tau+\varepsilon} \\ &+ \int_{\tau+\varepsilon}^{T} \left[\frac{\partial H_w^\varepsilon}{\partial f_w} \frac{\partial f_w^\varepsilon}{\partial \varepsilon} + \frac{\partial H_w^\varepsilon}{\partial p_w} \frac{\partial p_w^\varepsilon}{\partial \varepsilon} + \frac{d\lambda_2}{dt} \frac{\partial f_w^\varepsilon}{\partial \varepsilon} + \frac{d\mu_2}{dt} \frac{\partial p_w^\varepsilon}{\partial \varepsilon}\right] dt \\ &+ \left[\lambda_2 \frac{\partial f_w^\varepsilon}{\partial \varepsilon}\right]_{t=\tau+\varepsilon} - \left[\lambda_2 \frac{\partial f_w^\varepsilon}{\partial \varepsilon}\right]_{t=T} + \left[\mu_2 \frac{\partial p_w^\varepsilon}{\partial \varepsilon}\right]_{t=\tau+\varepsilon} - \left[\mu_2 \frac{\partial p_w^\varepsilon}{\partial \varepsilon}\right]_{t=T} \\ &+ \left[\frac{\partial \varphi^\varepsilon}{\partial f_w} \frac{\partial f_w^\varepsilon}{\partial \varepsilon} + \frac{\partial \varphi^\varepsilon}{\partial p_w} \frac{\partial p_w^\varepsilon}{\partial \varepsilon}\right]_{t=T} \end{aligned} \right\}_{\varepsilon=0}$$

## A2.4. Evaluating the change in the Lagrangian at zero

For $\tau$ to be optimal, the partial derivative of $L_\varepsilon$ must vanish (e.g. be equal to zero). Though it is certainly not the only way to satisfy [A10], the standard procedure is to group like terms in [A13] by their partial derivatives and set each of those components equal to zero (Conrad and Clark 1987).[2]

Grouping like terms in [A13] and setting it equal to zero,

[A14]
$$0 = \left.\frac{\partial L_\varepsilon}{\partial \varepsilon}\right|_{\varepsilon=0} = \left\{\begin{array}{l} \left[\left[c_c(f_c) + c_{p_c}(p_c)\right] - \left[c_w(f_w^\varepsilon) + c_{p_w}(p_w^\varepsilon)\right]\right]e^{-\delta t}\Big|_{t=\tau+\varepsilon} \\ + \int_{\tau+\varepsilon}^{T}\left[\left(\frac{\partial H_w^\varepsilon}{\partial f_w} + \frac{d\lambda_2}{dt}\right)\frac{\partial f_w^\varepsilon}{\partial \varepsilon} + \left(\frac{\partial H_w^\varepsilon}{\partial p_w} + \frac{d\mu_2}{dt}\right)\frac{\partial p_w^\varepsilon}{\partial \varepsilon}\right]dt \\ + \left[\left(\frac{\partial \varphi^\varepsilon}{\partial f_w} - \lambda_2\right)\frac{\partial f_w^\varepsilon}{\partial \varepsilon}\right]_{t=T} + \left[\lambda_2 \frac{\partial f_w^\varepsilon}{\partial \varepsilon}\right]_{t=\tau+\varepsilon} \\ + \left[\left(\frac{\partial \varphi^\varepsilon}{\partial p_w} - \mu_2\right)\frac{\partial p_w^\varepsilon}{\partial \varepsilon}\right]_{t=T} + \left[\mu_2 \frac{\partial p_w^\varepsilon}{\partial \varepsilon}\right]_{t=\tau+\varepsilon} \end{array}\right\}_{\varepsilon=0}$$

In order for the right-hand side of [A14] to vanish, the following conditions hold at an optimal reintroduction time:

$$\left[\left[c_c(f_c) + c_p(p_c)\right] - \left[c_w(f_w^\varepsilon) + c_p(p_w^\varepsilon)\right]\right]e^{-\delta t}\Big|_{t=\tau+\varepsilon}\Big|_{\varepsilon=0} = 0 \rightarrow$$

$$\left[\left[c_c(f_c) + c_p(p_c)\right] - \left[c_w(f_w) + c_p(p_w)\right]\right]e^{-\delta t}\Big|_{t=\tau} = 0$$

$$\left[\left(\frac{\partial H_w^\varepsilon}{\partial f_w} + \frac{d\lambda_2}{dt}\right)\frac{\partial f_w^\varepsilon}{\partial \varepsilon}\right]_{\varepsilon=0} = 0 \rightarrow \left(\frac{\partial H_w^\varepsilon}{\partial f_w} + \frac{d\lambda_2}{dt}\right)\Big|_{\varepsilon=0} = 0 \rightarrow \frac{\partial H_w}{\partial f_w} + \frac{d\lambda_2}{dt} = 0$$

$$\left[\left(\frac{\partial H_w^\varepsilon}{\partial p_w} + \frac{d\mu_2}{dt}\right)\frac{\partial p_w^\varepsilon}{\partial \varepsilon}\right]_{\varepsilon=0} = 0 \rightarrow \left(\frac{\partial H_w^\varepsilon}{\partial p_w} + \frac{d\mu_2}{dt}\right)\Big|_{\varepsilon=0} = 0 \rightarrow \frac{\partial H_w}{\partial p_w} + \frac{d\mu_2}{dt} = 0$$

$$\left[\lambda_2 \frac{\partial f_w^\varepsilon}{\partial \varepsilon}\right]_{t=\tau+\varepsilon}\Big|_{\varepsilon=0} = 0 \rightarrow \left[\lambda_2\right]_{t=\tau+\varepsilon}\Big|_{\varepsilon=0} = 0 \rightarrow \lambda_2\big|_{t=\tau} = 0$$

$$\left[\mu_2 \frac{\partial p_w^\varepsilon}{\partial \varepsilon}\right]_{t=\tau+\varepsilon}\Big|_{\varepsilon=0} = 0 \rightarrow \left[\mu_2\right]_{t=\tau+\varepsilon}\Big|_{\varepsilon=0} = 0 \rightarrow \mu_2\big|_{t=\tau} = 0$$

---

[2] As pointed out by a reviewer, grouping like terms in [A13] by their partial derivatives and setting the individual coefficients equal to zero is but one way of satisfying [A10]. Indeed, there may be other solutions that involve setting other groups of terms to zero, but this goes beyond the scope of this paper.

$$\left[\left(\frac{\partial \varphi^{\varepsilon}}{\partial f_w} - \lambda_2\right)\frac{\partial f_w^{\varepsilon}}{\partial \varepsilon}\right]\Bigg|_{t=T}\Bigg|_{\varepsilon=0} = 0 \to \left(\frac{\partial \varphi^{\varepsilon}}{\partial f_w} - \lambda_2\right)\Bigg|_{t=T}\Bigg|_{\varepsilon=0} = 0 \to \left(\frac{\partial \varphi}{\partial f_w} - \lambda_2\right)\Bigg|_{t=T} = 0$$

$$\left[\left(\frac{\partial \varphi^{\varepsilon}}{\partial p_w} - \mu_2\right)\frac{\partial p_w^{\varepsilon}}{\partial \varepsilon}\right]\Bigg|_{t=T}\Bigg|_{\varepsilon=0} = 0 \to \left(\frac{\partial \varphi^{\varepsilon}}{\partial p_w} - \mu_2\right)\Bigg|_{t=T}\Bigg|_{\varepsilon=0} = 0 \to \left(\frac{\partial \varphi}{\partial p_w} - \mu_2\right)\Bigg|_{t=T} = 0$$

Thus we may write the full suite of conditions for an optimal reintroduction time:

(the optimality condition or necessary first-order condition)
[A15] $\quad c_c(f_c(\tau)) + c_p(p_c(\tau)) = c_w(f_w(\tau)) + c_p(p_w(\tau)) \to c_c(f_c(\tau)) = c_w(f_w(\tau))$

(the adjoint equations)

[A16] $\quad\quad\quad\quad\quad \dfrac{d\lambda_2}{dt}(t) = -\dfrac{\partial}{\partial f_w} H_w(t, f_w(t), p_w(t))$

[A17] $\quad\quad\quad\quad\quad \dfrac{d\mu_2}{dt}(t) = -\dfrac{\partial}{\partial p_w} H_w(t, f_w(t), p_w(t))$

(initial and terminal time (transversality) conditions)
[A18] $\quad\quad\quad\quad\quad\quad\quad \lambda_2(\tau) = 0$
[A19] $\quad\quad\quad\quad\quad\quad\quad \mu_2(\tau) = 0$
[A20] $\quad\quad\quad\quad\quad\quad\quad \lambda_2(T) = \dfrac{\partial}{\partial f_w}\varphi(f_w(T), p_w(T))$
[A21] $\quad\quad\quad\quad\quad\quad\quad \mu_2(T) = \dfrac{\partial}{\partial p_w}\varphi(f_w(T), p_w(T))$

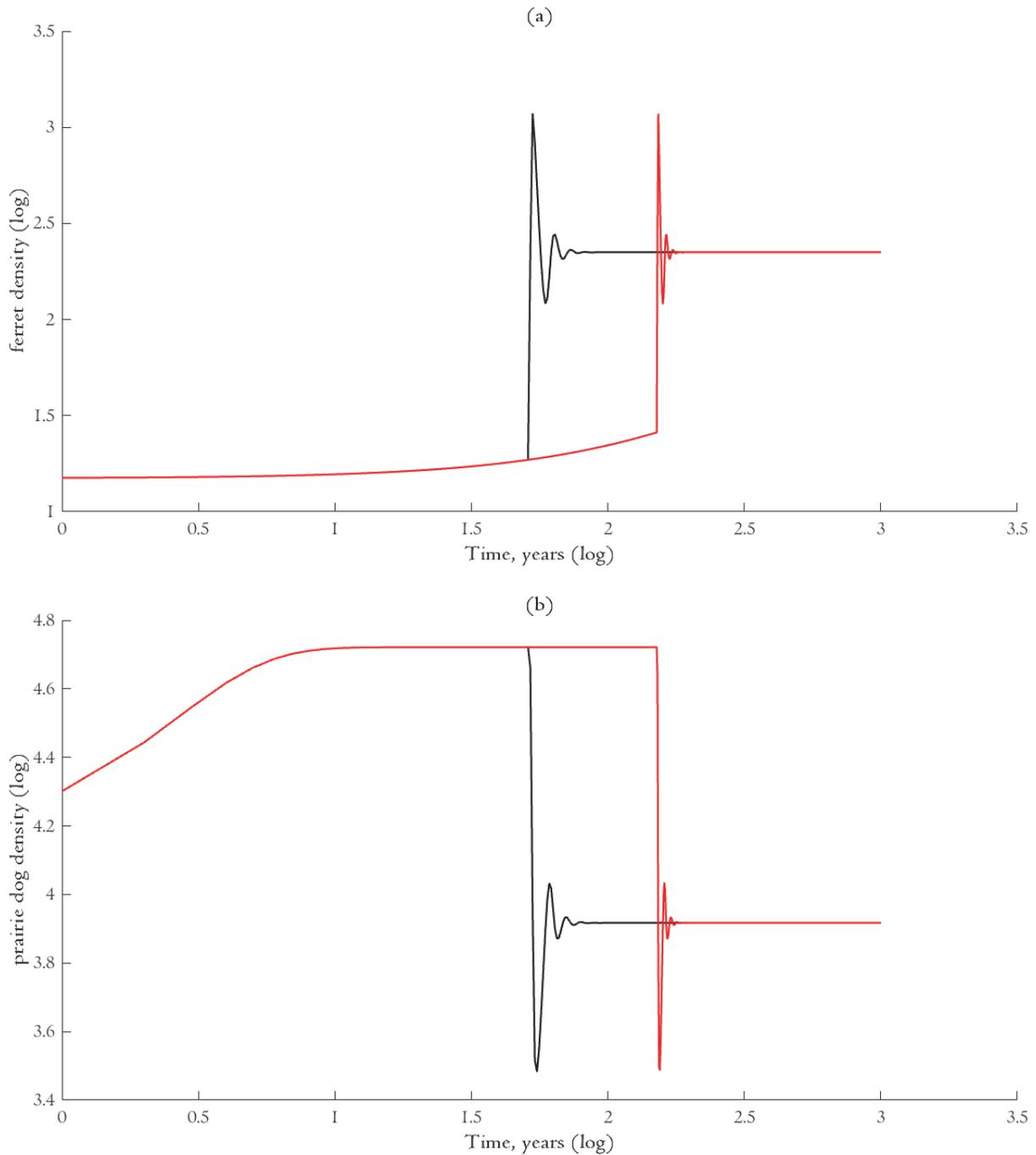

Figure S1. Effect of perturbing reintroduction time on population trajectories (orbits) of ferrets (a) and prairie dogs (b). Color indicates the reintroduction time: black ($\tau = 50$) and red ($\tau = 150$). While a longer reintroduction time changes the population size of each species at the time of reintroduction, it does not alter the trajectories of each species.

**Supplementary Material B. Deriving the optimal reintroduction time and ferret population size**

Substituting the economic functions specified in equations [13]-[15] into the necessary first order condition for optimization yields

[A22] $$l_c + a_c f_c + b_c f_c^{g_c} = a_w f_w$$

at $t = \tau^*$. If all captive ferrets are translocated into the wild at the reintroduction time and the convex costs of captive breeding can be approximated with a second-degree polynomial (i.e. $g_c = 2$), then Equation [A22] simplifies to

[A23] $$b_c f_c^2 + (a_c - a_w) f_c + l_c = 0$$

The solution to equation [A23] is $f_c^*$ - the size of the captive ferret population ($f_c$) at the optimal reintroduction time ($\tau^*$). The solution is restricted to real, positive values. Thus the equation for the optimal size of the captive ferret population at the time of reintroduction ($f_c^*$) is written,

[A24] $$f_c^* = \frac{-(a_c - a_w) + \sqrt{(a_c - a_w)^2 - 4 b_c l_c}}{2 b_c}$$

Assuming that the captive growth process is approximated with a linear term $a$ ($\dot{f}_c = a$) then a complete solution for the growth dynamics of captive ferrets is,

[A25] $$f_c(t) = at + f_c^0$$

where $f_c^0 = f_c(0)$ is the initial number of captive ferrets. With the complete solution in [A24] we can use equation [A25] to also derive the optimal reintroduction time ($\tau^*$),

$$f_c(t) = at + f_c^0 \rightarrow f_c^* = a\tau^* + f_c^0$$

[A26] $$\tau^* = \frac{f_c^* - f_c^0}{a}$$

**Supplementary Material C. Population dynamics and extinction**

Our primary focus is on the case in which ferrets and prairie dogs coexist in the wild, but the predator-prey system possesses corner solutions where one (ferrets) or both species die out. These are detailed in Table 2 and Figures S2 and S3.

      The model and parameters imply the existence of viable interior steady states for the three response functions. These steady states are described in Table 2. Phase plane diagrams are presented in Figure S3. The interior equilibria associated with the type I and type III response functions are stable spirals, and the interior equilibrium associated with the type II response function is an unstable spiral.[3] The type of response function thus has two important implications for ferret conservation. First, the equilibrium population of ferrets increases greatly from a type I to type II to type III responses, with the type III equilibria possessing more than double the number of ferrets than that of a type I response. This is due to the bounded predation rate in the type II and type III systems, particularly at low prairie dog biomasses (Figure 1a). Second, for the stable spirals the populations of prairie dogs and ferrets converge to the interior equilibrium, but for the unstable spiral the populations never stabilize and ferrets always go extinct unless the system starts at the steady state.

      Without further control over the system managers cannot alter the dynamics illustrated in the phase diagrams. After reintroduction, managers have little ability to influence the system and thereby change the stability landscape (Horan et al. 2011). This means that, unless the populations of ferrets and prairie dogs are just right by chance, for a type II functional response it will be impossible for the two species to coexist. Given these species coexisted prior to habitat development, and that their populations would have been subject to stochastic perturbations, we believe the type II response is empirically intractable, so we do not give it further consideration.

      In contrast, there is a large range of initial conditions under which the two species coexist for the type I and type III response functions. The associated simulation results are presented in Figure S2. Unless both populations are initially small relative to the steady state, then a system with the type III response will approach equilibrium significantly faster than a system with the type I response.

      Though unstable saddle points there are parts of the parameter space where the system crashes to the corner solutions. For a type I response this occurs at low levels of ferrets and prairie dogs (less than 3 ferrets, less than 101 prairie dogs) and at high populations of both species (Figure S3). This is due to the linear nature of the predator response function. For a type III response this occurs at low populations of either species (less than 10 ferrets, less than 500 prairie dogs). We were unable to achieve coexistence with the type II predator response. Given our parameter values, this is due to the unstable nature of the interior steady state solution.

---

[3] For Holling-type predator responses, the existence of a locally stable positive equilibrium implies global stability (Gotelli, 1995; May 1974; Kar and Matsuda, 2007; Kuang and Freedman, 1988). That is, knowledge of the locally stable population numbers for ferrets and prairie dogs at the terminal time, $T$, is sufficient to solve the management problem from an optimal transition time, $\tau^*$.

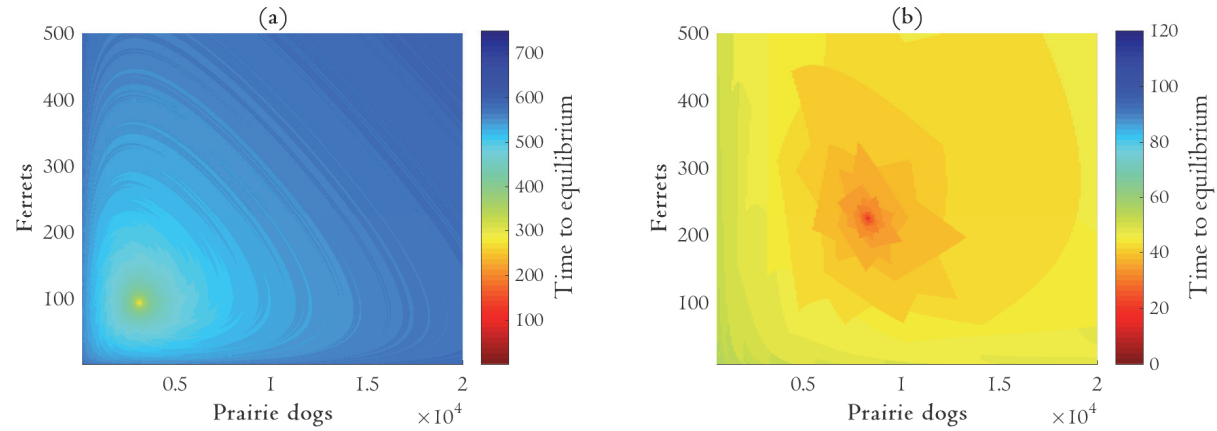

Figure S2. Coexistence of ferrets and prairie dogs under varying initial conditions for type I (a) and type III (b) functional responses. Color indicates the number of time steps required for both species to reach $\epsilon$ percent from the equilibrium value. Note the difference in scales for the time to equilibrium of type I and type III responses.

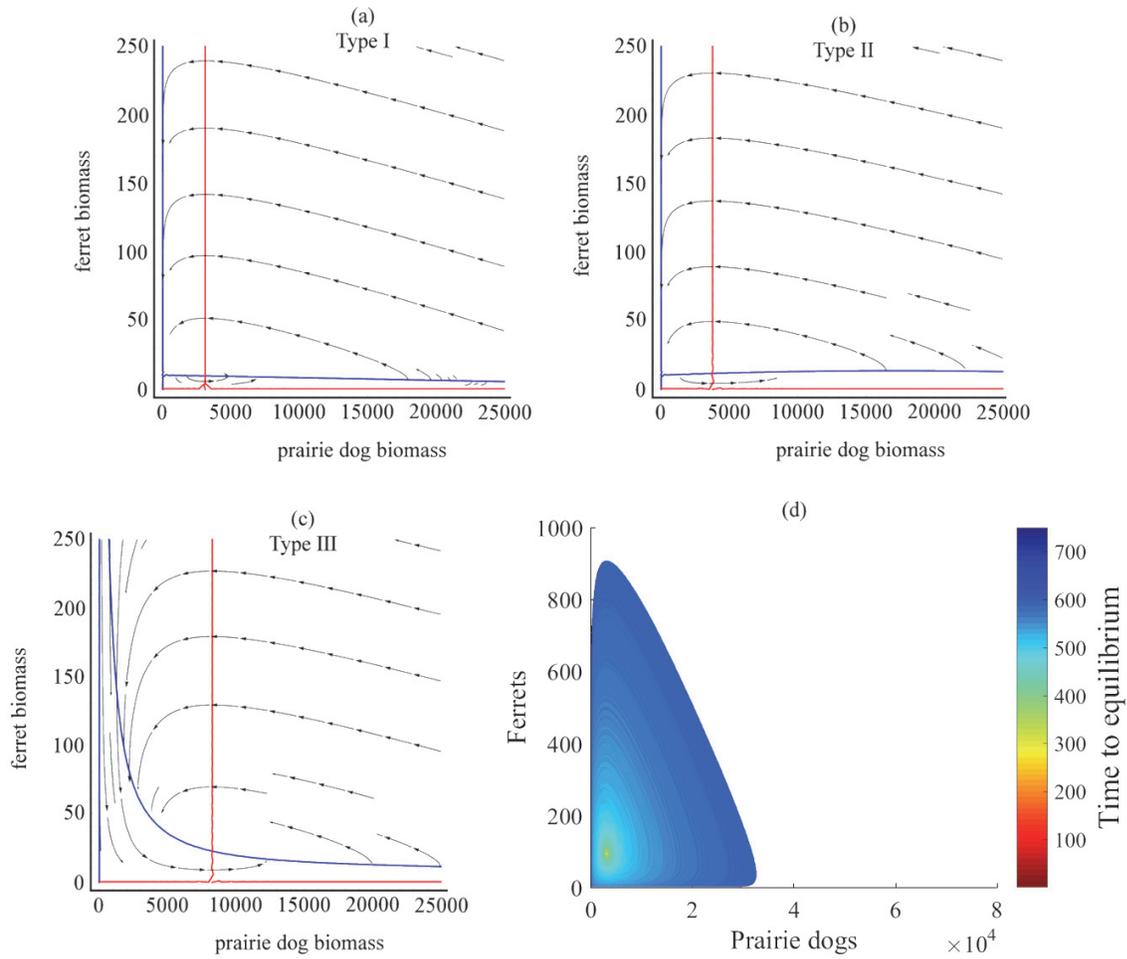

Figure S3. Phase plane diagrams for type I-III predator responses (a-c) and regions of species coexistence (d, type I predator response). (a-c) Color denotes the prairie dog (blue) and ferret (red) zero-isoclines. (d) Color indicates the number of time steps required for both species to reach $\epsilon$ percent from the equilibrium value. White indicates areas of the parameter space where coexistence is not possible.

**Supplementary Material D. Derivation of captive ferret growth rate**

      Captive ferret growth rate was calibrated from data from the Phoenix Zoo captive breeding program (Figure S4). A large number of ferrets are exchanged annually between national captive breeding sites. Therefore, we approximated growth rate by regressing the number of ferrets on-site at the end of the calendar year against time. This yielded the average increase in the local ferret population per year (2.776 ferrets per year, $P < 0.008$, $R^2 = 0.60$), which we converted to the annual average percent increase (0.0752%). The discrete-time multiplier for captive ferret growth rate was 1.075. We calculated the continuous-time equivalent as:

$$e^a = 1.075 \rightarrow a = \ln(1.075) = 0.0725$$

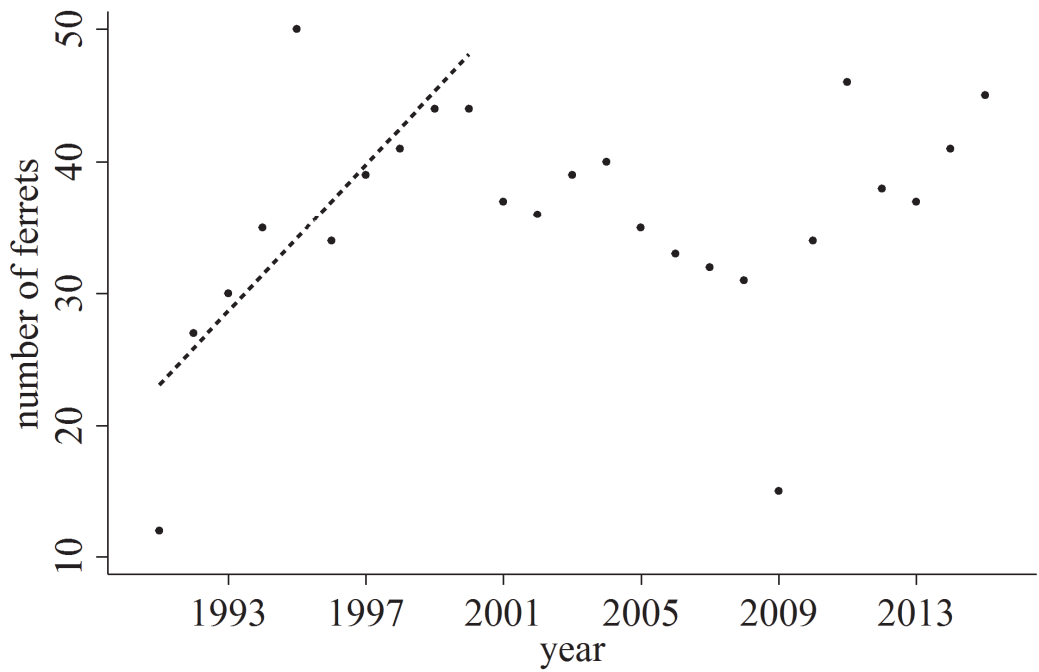

Figure S4. Population census data for the Phoenix Zoo captive breeding program. The interval between 1991 and 2000 provided the best linear fit with the data (dashed line; $R^2 = 0.60$). Therefore, we focused on that time period for our analysis.